\documentclass[aps,prd,twocolumn,showpacs,nofootinbib]{revtex4}

\usepackage{amssymb}

\newcommand{\h}[1]{{}^{\mbox{\,\tiny $\{#1\}\!$}}h}
\newcommand{\pert}[2]{{}^{\mbox{\,\tiny $\{#1\}\!$}}{#2}}

\newcommand{\E}[7]{{E_{#1}^{#4}{}_{#2}^{#5}{}_{#3}^{#6}
{}^{}_{#7}}}

\newcommand{\onlyU}[1]{{{}^{\mbox{\tiny $({#1})\!$}}{\cal U}}}
\newcommand{\U}[7]{{{}^{\mbox{\tiny $({#1})\!$}}{\cal U}_{#2}^{#3}
{}_{#4}^{#5}{}_{#6}^{#7}}}
\newcommand{\Es}[7]{{{}^{\mbox{\tiny $({#1})\!$}}{\cal E}_{#2}^{#3}
{}_{#4}^{#5}{}_{#6}^{#7}}}
\newcommand{\Estilde}[7]{{{}^{\mbox{\tiny $({#1})\!$}}{\tilde{\cal E}}_{#2}^{#3}
{}_{#4}^{#5}{}_{#6}^{#7}}}
\newcommand{\Source}[7]{{{}^{\mbox{\tiny $({#1})\!$}}S_{#2}^{#3}
{}_{#4}^{#5}{}_{#6}^{#7}}}

\begin{document}

\title{High-order perturbations of a spherical collapsing star}

\pacs{04.25.Nx, 04.30.Db, 04.40.Dg}

\author{David Brizuela}
\affiliation{Theoretisch-Physikalisches Institut, Friedrich-Schiller-Universit\"at,
Max-Wien-Platz 1, 07743 Jena, Germany}
\affiliation{Instituto de Estructura de la Materia, CSIC, Serrano
121-123, 28006 Madrid, Spain}
\author{Jos\'e M. Mart\'in-Garc\'ia}
\affiliation{
Institut d'Astrophysique de Paris, Univ. Pierre et Marie Curie,
CNRS, 98bis boul. Arago, 75014 Paris, France}
\affiliation{Laboratoire Univers et Th\'eories, CNRS,
Univ. Paris Diderot, 5 place Jules Janssen, 92190 Meudon, France}
\affiliation{Instituto de Estructura de la Materia, CSIC, Serrano
121-123, 28006 Madrid, Spain}
\author{Ulrich Sperhake}
\affiliation{Institut de Ci\`encies de l'Espai (CSIC-IEEC), Campus UAB, E-08193 Bellaterra, Spain}
\affiliation{Theoretical Astrophysics 350-17, California Institute of Technology, Pasadena, CA 91125}
\affiliation{Department of Physics and Astronomy, The University of Mississippi, University, MS 38677, USA}
\author{Kostas D. Kokkotas}
\affiliation{Theoretical Astrophysics, Eberhard-Karls University of T\"ubingen, 72076 T\"ubingen, Germany}

\begin{abstract}
In Ref. \cite{BMM06, BMM07} a formalism to deal with high-order
perturbations of a general spherical background was developed.
In this article, we apply it to the particular case of a perfect
fluid background. We have expressed the perturbations of the
energy-momentum tensor at any order in terms of the perturbed
fluid's pressure, density and velocity. In general, these expressions are
not linear and have sources depending on lower order perturbations.
For the second-order case we make the explicit decomposition of these
sources in tensor spherical harmonics. Then, a general procedure is
given to evolve the perturbative equations of motions of the perfect
fluid for any value of the harmonic label. Finally, with the problem of a
spherical collapsing star in mind, we discuss the high-order perturbative
matching conditions across a timelike surface, in particular the surface
separating the perfect fluid interior from the exterior vacuum.
\end{abstract}

\maketitle

\section{Introduction}

The detection of gravitational waves is currently considered to be one
of the most important open problems in astrophysics and experimental
physics. Aside from
providing a test of the theory of general relativity in the strong-field
regime, it will open a new window for astrophysical
observations, giving rise to the era of {\em gravitational-wave astronomy}.
To this end the ground-based laser-interferometric detectors
LIGO, VIRGO, GEO600 and TAMA \cite{Abbott2007a, Beauville2007, Lueck2006,
Tatsumi2007} are now collecting data at or near design sensitivity,
while next generation ground-based detectors like the Einstein
Telescope (ET) are in the research and development phase
\cite{Punturo10}.
The space based interferometer LISA \cite{Danzmann2003}, scheduled for
launch in the next decade, will complement such observations with
high precision in the low-frequency range. Gravitational wave
detection and interpretation
requires a close interplay between the experimental side and theoretical
predictions as physical signals need to be dug out from a noisy
data stream. Various astrophysical scenarios are expected to generate
gravitational radiation that will be detectable on Earth. Most
prominently among these feature the inspiral and coalescence
of compact binaries, black holes and/or neutron stars, as well as
stellar collapse. The work of this article is primarily addressed
at the modelling of the latter type of sources although we believe
our results to be of interest for a variety of questions pertaining
to gravitational wave source modeling.

Simulating neutron stars and black holes in the framework
of fully non-linear general relativity is only possible using numerical
methods on super computers, an area of research commonly referred to as
{\em numerical relativity}. This field has made enormous progress recently;
see, for example, the reviews \cite{Fon03, Pretorius2007a, Ott2008}.
In spite of this progress,
approximative techniques such as the post-Newtonian expansion and
perturbation theory still play a crucial role in the modeling of
astrophysical sources of gravitational waves. First, numerical simulations
are computationally too expensive to cover more than a few tens of
orbits in inspirals, so that complete waveforms are now constructed
by hybridization of numerical and post-Newtonian results
\cite{Ajith2007, Ajith2007a} or calibration of the effective-one-body
model via numerical relativity simulations
\cite{Damour2008, Buonanno2009, Pan2010}. Second, other
types of sources completely elude a full numerical treatment. For
example, simulations of the so-called
{\em extreme mass ratio binaries} are practical only in
a perturbative framework; see \cite{Poisson2004, Hinderer2008} and
references therein for a discussion
as well as \cite{Gonzalez2008, Lousto:2010tb, Lousto:2010qx, Lousto:2010ut}
for the most
extreme mass ratios simulated in numerical relativity.
Perturbation theory
also provides a natural framework for interpreting numerical results,
as for example in the case of black-hole and neutron-star
oscillations \cite{Kokkotas1999, Berti2009}.
Finally perturbative calculations
efficiently provide results with high precision even when involving
numerical methods and are thus ideally suited for the study of
a variety of astrophysical and cosmological problems as well as
benchmarking the accuracy of full blown numerical simulations.

Purpose of this work is to provide a general gauge invariant framework
for perturbation theory of arbitrary order on time-dependent and
spherically-symmetric perfect fluid backgrounds. This work constitutes
a particularization of the formalism developed in Refs.~\cite{BMM06, BMM07}
by including matter obeying perfect fluid equations of state.
In particular, we will also include a careful analysis
of the matching conditions necessary to obtain a consistent treatment
of the stellar interior and the vacuum exterior spacetime.

Static backgrounds represent a particularly simple subclass
of configurations because they allow for a Fourier expansion of the
perturbations in such a way that reduces the problem to a set of
ordinary differential equations depending on radius only.
Examples for the general type of background configurations
considered in this work include radially collapsing or exploding stars.
The collapse of homogeneous dust spheres studied extensively in a
series of papers by Cunningham, Price and
Moncrief~\cite{CPM78, CPM79, CPM80} can be viewed as the special case
of a perfect fluid with vanishing pressure. In this case
the different layers of the star do not interact which leads to a
simpler type of motion but may result in unphysical consequences.

The perturbations can be either radial or non-radial,
although in some scenarios it turns out to be convenient to
absorb radial perturbations in the
spherically symmetric background configuration. See, for example,
Ref.~\cite{Gabler2009} for a fully non-linear numerical treatment of
radial pulsations of neutron stars. Non-linearities are
naturally incorporated in perturbation theory by including higher-order
terms in the calculation. First-order perturbation theory,
for example, linearizes the Einstein equations in the first-order
metric perturbations $\pert{1}{h}_{\mu\nu}$ whereas second-order
theory results in a set of differential equations linear in the
second-order metric perturbations $\pert{2}{h}_{\mu\nu}$
but containing source terms quadratic in the $\pert{1}{h}_{\mu\nu}$.
This hierarchical structure can be extended to arbitrary order
although the equations quickly become overwhelmingly complicated.

The application of perturbation theory to compact stars in a general
relativistic framework was
pioneered in the 1960s by Chandrasekhar's analysis of the stability
of static-background models against radial perturbations
\cite{Cha64, Cha64*}. Next, Thorne and collaborators established the
theoretical basis for the study of non-radial perturbations of
perfect fluid stars \cite{ThCa67, PrTh69, Tho68, CaTh70, IpTh73}.
Further development and applications can be found
in Refs.~\cite{ChFe91,IpPr90,KoSc92, AnKo96}.
Time-dependent background configurations were first studied
in the above mentioned work by Cuningham et al.~\cite{CPM78, CPM79}
on collapsing dust spheres.
Seidel and coworkers \cite{SeMo87, SMM88, Sei90} applied the formalism
of Gerlach and Sengupta \cite{GeSe79, GeSe80}
to generalize the work of Cunningham et al. to general perfect fluid
equations of state. Motivated by this work, Gundlach and
Mart\'in-Garc\'ia \cite{GuMa00} developed a covariant and gauge-invariant
framework to analyze arbitrary first-order perturbations of a spherical
perfect fluid. This framework was later used by Harada et al. \cite{HIS03}
to analyze axial perturbations of stellar collapse.

To our knowledge higher-order perturbation theory of stellar models
has so far only been applied to the modeling of slowly, uniformly
rotating stars. The key idea here is to incorporate the rotation
in the form of a first-order axial perturbation. This idea has again
been pioneered in the Cunningham et al. series \cite{CPM80}
in the special case of a collapsing dust sphere. Applications to
a wider class of perfect fluids have so far been restricted to
static background models \cite{Koj92, ChFe91*}. Differential
rotation has been modelled in the context of the
Cowling approximation\footnote{In this approximation one
neglects metric perturbations and thus evolves fluid
perturbations only.} \cite{SPK07, PSK07}. More recently it has
become possible to study perturbations of fast rotating axisymmetric
configurations, still in the Cowling approximation
\cite{GK08,KGK10,GK10}.

Our work represents the natural extension of the existing literature by
constructing a general framework for second-order
perturbations on a time-dependent, spherically symmetric background
spacetime containing matter in the form of a perfect fluid star.
The starting point for this construction is given by the second-order
generalization of the Gerlach-Sengupta formalism developed in
Refs.~\cite{BMM06, BMM07}. Specifically, we will apply this gauge-invariant
formalism to the case of a perfect fluid star following
the notation and techniques presented in Ref.~\cite{GuMa00}.

A key ingredient in our study is the matching at second
perturbative order of the stellar interior and the vacuum exterior
parts of the spacetime. Its first order analogue was analyzed in
a covariant and gauge invariant manner in \cite{MaGu01}, but we will
see that the extension to higher orders represents a
non-trivial problem. Its solution requires a careful analysis as to
which quantities must be continuous across the stellar surface.
Our analysis of this particular point will be valid at any
perturbative order
and we believe, it will also be helpful in clarifying the
matching conditions in certain background configurations.

A brief overview of the quantities and the equations governing their
behavior is given as follows. The background metric in the stellar
interior is described by the scalars $U$ and $W$ defined in
Eq.~(\ref{eq:defUV}) and the structure coefficients $\mu$ and
$\nu$ given in Eq.~(\ref{eq:defmunu}), all defined
with respect to the background velocity field. The fluid is described
by its density $\rho$, the entropy per fluid element $s$ and
the pressure $p$. These variables are determined by
Eqs.~(\ref{Uprime})-(\ref{sdot}) and the equation of state
that provides $p$ as a function of $\rho$ and $s$.

The perturbations at order $n$ are naturally divided into axial and
polar modes and further decomposed into multipoles with labels $l$
and $m$. For simplicity, we will ommit these labels from the perturbation
variables in this summary as is also done below in the derivation.
The axial perturbations are given by
the velocity component $\beta$ (\ref{eq: d2u}) and the Gerlach-Sengupta
master scalar $\Pi$ (\ref{Pidefinition}) which are evolved according to
Eqs.~(\ref{master}) and (\ref{matteraxial}). The metric components
$\delta$ and $\lambda$ (\ref{framevector}) can
be reconstructed from the master function $\Pi$ according to
Eqs.~(\ref{reconstruct1}), (\ref{reconstruct2}). The polar
perturbations are given by the velocity components
$\alpha$ and $\gamma$ (\ref{eq: d2u}) the
energy and entropy perturbations $\omega$ and $\sigma$
(\ref{defomega}), (\ref{defsigma}), the metric perturbations
$\eta$, $\chi$ and $\psi$ defined in Eqs.~(\ref{dectensor}), (\ref{defchi})
and the scalar $\mathcal{K}$ (\ref{eq: pert_metric}).
This set of perturbations obeys the evolution system
(\ref{fluideq7bis}), (\ref{polareq1})-(\ref{polareq6}), (\ref{polareq10}).

The remainder of this article is organized as follows.
Section \ref{sec: hogs} presents a brief summary of the generalized
formalism introduced in Ref. \cite{BMM06, BMM07}. In
Sec.~\ref{sec: perfect_fluid} we introduce the notation for
the background fluid variables and their perturbations.
We further express the perturbation of the energy-momentum tensor in
terms of these variables. The evolution equations resulting from the
perturbed Einstein and matter equations
are presented in Sec.~\ref{sec: evol_eqs}. The high-order matching
conditions are the subject of Sec.~\ref{sec: matching}. This section
is quite independent from the previous ones, so the
interested reader is referred directly there. We conclude in
Sec.~\ref{sec: conclusions} with a discussion of future applications
and extensions of our work.

The intensive tensor computations in this work have been performed with
the Tensor Computer Algebra framework {\em xAct} \cite{xAct}, based on
{\em Mathematica}. Of particular importance has been the package
{\em xPert} \cite{xPert} for high-order metric perturbation theory
around curved backgrounds.

\section{High-order Gerlach and Sengupta formalism}
\label{sec: hogs}

\subsection{Background spherical spacetime}

The spherically symmetric nature of the background spacetime enables
us to naturally decompose it as
${\cal{M}}\equiv {\cal{M}}^2\times S^2$, where ${\cal M}^2$
is a two-dimensional Lorentzian manifold and $S^2$ the
unit two-sphere.
In order to distinguish tensor fields residing on these different manifolds,
we will use Greek letters $(\mu,\nu,\dots)$ for four-dimensional indices,
capital Latin letters $(A,B,\dots)$ for indices on ${\cal M}^2$ and
lower-case Latin letters $(a, b,\dots)$ for indices on the sphere.
With this notation, we can decompose
the background metric and energy-momentum tensor as
\begin{eqnarray} \label{sphericalgdecomposition}
  g_{\mu\nu}(x^D, x^d)dx^\mu dx^\nu &=& g_{AB}(x^D)dx^A dx^B
\nonumber\\ &+&
  r^2(x^D)\, \gamma_{ab}(x^d)dx^a dx^b ,
\\ \label{sphericaltdecomposition}
  t_{\mu\nu}(x^D, x^d)dx^\mu dx^\nu &=& t_{AB}(x^D)dx^A dx^B
\nonumber\\  &+&
 \frac{1}{2} r^2(x^D) Q(x^D)\gamma_{ab}(x^d)dx^a dx^b \!,\quad
\end{eqnarray}
where $g_{AB}$ is the metric on ${\cal M}^2$, $\gamma_{ab}$ is the standard
metric on the unit sphere and $r,Q$ are scalar fields on ${\cal M}^2$.
We denote covariant derivatives associated with the different metrics by
\begin{equation}
  g_{\mu\nu;\lambda}=0,\quad g_{AB|C}=0,\quad \gamma_{ab:c}=0,
\end{equation}
and define the vector $v_A\equiv r_{|A}/r$.
The Einstein equations for the metric
(\ref{sphericalgdecomposition}) with matter tensor
(\ref{sphericaltdecomposition}) are given in \cite{GeSe79}.

Tensor fields of any rank $s$ on the sphere will be decomposed using 
a basis of tensor spherical harmonics labeled by
multipole indices $l$ and $m$. Such basis can be constructed from
the symmetric trace-free (STF) tensors
\begin{eqnarray}
  Z_l^m{}_{a_1...a_s}&\equiv &\left(Y_l^m{}_{:a_1...a_s}\right)^{\rm STF},
      \label{eq: Zlm} \\\label{Xlm}
  X_l^m{}_{a_1...a_s}&\equiv &\epsilon_{(a_1}{}^b Z_l^m{}_{ba_2...a_s)},
\end{eqnarray}
which are respectively polar and axial,
together with the metric $\gamma_{ab}$ and the antisymmetric tensor
$\epsilon_{ab}$ \cite{BMM06}. For the particular case $s=0$, those
objects must be read as $Z_l^m\equiv Y_l^m$ and $X_l^m\equiv 0$.

At first order modes with different $l,m$ and polarities decouple.
However, at second and higher perturbative orders, the
non-linear coupling between first-order modes results
in products of tensor spherical harmonics with different labels
$(\hat l,\hat m,\hat s)$ and $(\bar l, \bar m, \bar s)$.
Those products can be decomposed into a linear combination of harmonics
$(l,\hat m+\bar m,\hat s+\bar s)$
with an explicit formula involving coefficients
\begin{equation}
\E{\bar s}{\bar l}{\bar m}{\hat s}{\hat l}{\hat m}{l}\propto
C_{\bar l}^{\bar m}{}_{\hat l}^{\hat m}{}_{l}^{\hat m+\bar m} C_{\bar l}^{\bar s}{}_{\hat l}^{\hat s}{}_{l}^{\hat s+\bar s},
\end{equation}
where $C$ stands for Clebsch-Gordan coefficients.
These $E$-coefficients encode the geometric selection rules that determine
which pairs of modes do actually couple. See \cite{BMM06} for full details.

\subsection{Non-spherical perturbations}

Perturbative calculations are often formulated in terms
of a one-parameter family of spacetimes
$({\cal M}(\varepsilon),g(\varepsilon))$. The expansion parameter
represents a measure for how strongly a physical system deviates from
the background configuration corresponding to $\varepsilon=0$.
In complete analogy,
any tensorial quantity of the system is represented by a
family $\Omega(\varepsilon)$, and expanded around
its background value $\Omega\equiv\Omega(\varepsilon=0)$
according to
\begin{equation}
  \Omega(\varepsilon)=\Omega+\sum_{n=1}^{\infty}
       \varepsilon^n\frac{\Delta^n[\Omega]}{n!}.
\end{equation}
A key feature of this expansion is that
the perturbations $\Delta^n[\Omega]$, also denoted by $\pert{n}{\Omega}$
in this work,
are tensors on the background manifold. We have thus converted a
single problem on an unknown manifold into an infinite hierarchy of
problems on the chosen background manifold.
There remains, however, one important problem;
we need to specify a diffeomorphism
$\phi(\varepsilon)$ relating points on the background ${\cal M}$
and the perturbed manifolds ${\cal M}(\varepsilon)$. The actual values
of the perturbations $\pert{n}{\Omega}$ depend on this choice of
{\em gauge}.

We next explicitly decompose tensorial perturbations in terms of the
basis of tensor harmonics introduced in Eqs.~(\ref{eq: Zlm}) and
(\ref{Xlm}). In this expansion the metric perturbations are
given by,
\begin{widetext}
\begin{eqnarray}\label{metricdecomposition}
  \Delta^n[g_{\mu\nu}] \equiv \h{n}_{\mu\nu} \equiv \sum_{l,m}
  \left(
      \begin{array}{cc}
        \pert{n}{H}{}_l^m\!{}_{AB} \; Z_l^m
        &
        \pert{n}{H}{}_l^m\!{}_A \; Z_l^m{}_b+
        \pert{n}{h}{}_l^m\!{}_A \; X_l^m{}_b
        \\
        \pert{n}{H}{}_l^m\!{}_B \; Z_l^m{}_a +\pert{n}{h} {}_l^m\!{}_B \;
        X_l^m{}_a
        &
        \;\pert{n}{K}_l^m \; r^2\gamma_{ab} Z_l^m +
        \pert{n}{G}_l^m \; r^2Z_l^m\!{}_{ab} + \pert{n}{h}_l^m \;
        X_l^m\!{}_{ab}
      \end{array}
  \right).
\end{eqnarray}
\end{widetext}
At each perturbative order $n$, the gauge freedom enables us to choose
freely four of these functions, three polar and one axial.
Alternatively to thus fixing the gauge, we can construct
{\em gauge-invariant} combinations of the perturbation functions
and work with those quantities.
We will follow this latter approach, though with a construction based
on the choice of a particular gauge.

For this purpose we consider the most natural
gauge choice in spherical symmetry:
Regge-Wheeler (RW) gauge \cite{ReWh57}.
It is defined by setting
\begin{equation}\label{RW}
  \pert{n}{H}{}_l^m\!{}_A=0,\quad\pert{n}{G}_l^m=0,\quad
  \pert{n}{h}_l^m=0,
\end{equation}
for all $n\ge 1$ and $l\ge 2$. We note that this does not constitute
a {\em rigid} choice, that is, there remains some gauge freedom
in the $l=0$ and $l=1$ modes.
Next we use an important result from Ref.~\cite{BMM07},
namely a procedure to define
gauge-invariant quantities whose values coincide with those of
the perturbations in RW gauge. The key idea of this
procedure is to employ the formula for general gauge transformations
parametrized by the generators $\pert{n}{\xi}$, changing the
perturbation $\Delta^n[\Omega]$ to
\begin{eqnarray}\label{ngaugetrans}\nonumber
  \Delta^n[\Omega] &+& \sum_{m=1}^{n}\frac{n!}{(n-m)!}\sum_{(K_m)}
                      \frac{1}{2!^{k_2}...(m!)^{k_m}k_1!...k_m!}\\
                   &\times&\quad{\mathcal L}_{\pert{1}{\xi}}^{k_1}...
                      {\mathcal L}_{\pert{m}{\xi}}^{k_m}
                      \Delta^{n-m}[\Omega],
\end{eqnarray}
where the second sum extends to the $m$-tuples
\begin{equation}
   (K_m)=\left\{(k_1,...,k_m)
         \in\mathbb{N}_0^m; \quad \sum_{j=1}^{m}j
         k_j=m\right\},
\end{equation}
$\mathbb{N}_0$ being the set of non-negative integers.
We then need to choose the vectors $\pert{m}{\xi}$ such that they take the
metric perturbation (\ref{metricdecomposition}) from an arbitrary gauge
to the RW form (\ref{RW}). Naturally, the vectors $\pert{m}{\xi}$ depend
on the perturbations $\pert{k}{h}_{\mu\nu}$ themselves, and so the
expressions for the gauge invariants are highly nontrivial beyond first
order.  Explicit formulas at second order, already expanded in terms of
the harmonic coefficients, are given in \cite{BMM07}.

Given that the perturbations in RW gauge and the associated
gauge invariants coincide in value, we will perform all our computations
in RW gauge, for simplicity, and then use the expressions in \cite{BMM07}
whenever we want to generalize the results to an arbitrary gauge.

In this way, the decomposition of the perturbations of the metric
and the energy-momentum tensor will be as follows,
\begin{widetext}
\begin{eqnarray}
  \h{n}_{\mu\nu} \equiv \sum_{l,m} \left(
      \begin{array}{cc}
        \pert{n}{\cal K}{}_l^m\!{}_{AB} \; Z_l^m &
        \pert{n}{\kappa}{}_l^m\!{}_A \; X_l^m{}_b \\
        \pert{n}{\kappa} {}_l^m\!{}_B \; X_l^m{}_a &
        \pert{n}{\cal K}_l^m \; r^2\gamma_{ab} Z_l^m
      \end{array}
  \right),
  \label{eq: pert_metric}
\end{eqnarray}
\begin{eqnarray}\label{emdecomposition}
  \Delta^n[t_{\mu\nu}] \equiv
  \sum_{l,m}
  \left(
      \begin{array}{cc}
        \pert{n}{\Psi}{}_l^m\!{}_{AB} \; Z_l^m &
        \pert{n}{\Psi}{}_l^m\!{}_A \; Z_l^m{}_b
        +\pert{n}{\psi}{}_l^m\!{}_A \; X_l^m{}_b \\
        \pert{n}{\Psi}{}_l^m\!{}_B \; Z_l^m{}_a
        +\pert{n}{\psi}{}_l^m\!{}_B \; X_l^m{}_a &
        \pert{n}{\tilde\Psi}_l^m \; r^2\gamma_{ab} Z_l^m
        +\pert{n}{\Psi}_l^m \; Z_l^m\!{}_{ab}
        +\pert{n}{\psi}_l^m \; X_l^m\!{}_{ab}
      \end{array}
  \right).
\end{eqnarray}
\end{widetext}
Note that polar (axial) harmonic coefficients are represented by
upper-case (lower-case) letters.
In order to agree with the construction by Gerlach and Sengupta
\cite{GeSe79, GeSe80},
we combine the invariants $\pert{n}{\psi}^m_l\!{}_A$ and
$\pert{n}{\kappa}^m_l\!{}_A$
to form a new matter invariant quantity
\begin{equation}\label{newmatterinvariant}
\pert{n}{\tilde\psi}^m_l\!{}_A
\equiv
\pert{n}{\psi}^m_l\!{}_A-\frac{Q}{2}\pert{n}{\kappa}^m_l\!{}_A.
\end{equation}

\section{Perfect fluid}
\label{sec: perfect_fluid}

The decomposition of metric and matter perturbations given above
is valid for arbitrary types of matter sources. In this section
we discuss in detail the special case of a perfect fluid.
We first formulate the equations for the background spacetime
in Sec.~\ref{sec: perfect_fluid_bg}, following \cite{GuMa00}.
Then we address the description of fluid perturbations in
Sec.~\ref{sec: perfect_fluid_pert}, generalizing the notations
and results in \cite{GuMa00} from first to higher orders.
The metric and fluid perturbative evolution equations will
be analyzed in the next section.

\subsection{Background perfect fluid}
\label{sec: perfect_fluid_bg}

The energy-momentum tensor for a perfect fluid with four velocity $u^\mu$,
total enery density $\rho$ and pressure $p$ is
\begin{equation}\label{fluidemtensor}
  t_{\mu\nu}\equiv (\rho + p)u_\mu u_\nu + p g_{\mu\nu}.
\end{equation}
This corresponds to the special case $Q=2p$ in the general expression
(\ref{sphericaltdecomposition}). We will use a generic equation of
state for a perfect fluid, in the form
$p=p(\rho,s)$ where $s$ is the entropy per particle. It is convenient
to introduce the partial derivatives of the pressure as additional
quantities according to
\begin{equation}\label{thermo}
  c_s^2\equiv \left(\frac{\partial p}{\partial\rho}\right)_s,
  \qquad C\equiv \frac{1}{\rho}\left(\frac{\partial p}{\partial s}\right)_\rho.
\end{equation}
Here, $c_s^2$ is the adiabatic
speed of sound, but we are not aware of a comparatively
intuitive meaning of the second quantity.

In spherical symmetry the four-velocity of the fluid takes the form
$u^\mu=(u^A,0)$. This vector defines a unique outward pointing spacelike
unit vector $n_A \equiv  -\epsilon_{AB} u^B$ on ${\cal M}^2$,
with $\epsilon_{AB}$ being the antisymmetric volume form on ${\cal M}^2$.
These two vectors form an orthonormal basis on this manifold,
which can be used to decompose all geometrical objects, for example
\begin{equation}
  g_{AB} = - u_A u_B + n_A n_B,
  \qquad \epsilon_{AB} = n_A u_B - u_A n_B.
\end{equation}
Using this decomposition, the ${\cal M}^2$ part of the energy-momentum
tensor (\ref{sphericaltdecomposition}) takes the form
\begin{equation}
  t_{AB} =  \rho u_A u_B + p n_A n_B.
\end{equation}
For scalar fields $f$ on ${\cal M}^2$ we further define the
{\em frame derivatives}
\begin{equation}
  \dot f \equiv  u^A f_{|A}, \qquad f' \equiv  n^A f_{|A}.
\end{equation}
A straightforward calculation shows that the frame derivatives obey
the commutation relation
\begin{equation}
\label{commutator}
  (\dot f)'-(f')\dot{}=\mu f' - \nu \dot f,
\end{equation}
where
\begin {equation}
  \mu\equiv {u^A}_{|A}, \qquad \nu\equiv {n^A}_{|A}.
  \label{eq:defmunu}
\end{equation}
These ``structure" functions are the components of the
covariant derivatives of the frame vectors in their own frame
\begin{equation}
  u_{A|B}=n_A (n_B \mu - u_B\nu), \quad
  n_{A|B}=u_A(n_B \mu - u_B\nu).
\end{equation}

We are now in the position to formulate the Einstein equations
exclusively in terms of the background scalars
$\rho$, $p$, $\mu$, $\nu$ and
\begin{equation}
  U\equiv u^A v_A, \qquad W\equiv n^A v_A.
  \label{eq:defUV}
\end{equation}
For our spherically symmetric background we thus obtain
expressions with remarkable symmetry for the four
independent field equations
\begin{eqnarray}\label{Uprime}
  U' && = W(\mu - U),\\
  \dot W && = U(\nu - W),\\
  W' && = - 4\pi \rho - W^2 + U\mu + {M\over r^3},\\
  \dot U && = - 4\pi p - U^2 + W\nu - {M\over r^3},
\end{eqnarray}
with the Hawking mass \cite{Haw} in spherical symmetry
(Misner-Sharp mass)
\begin{equation}\label{Hawmass}
  M\equiv \frac{r}{2}\left(1-r_{,A}r^{,A}\right)
       = \frac{r}{2}\left[1+r^2(U^2-W^2)\right].
\end{equation}

We still need to consider the equations of motion for the
perfect fluid. Conservation of energy-momentum results in
\begin{eqnarray}
  \dot\rho + \left(\rho+p\right)(2U+\mu) & = & 0, \\\label{Euler}
  c_s^2 \rho' + \rho C s' + \left(\rho+p\right) \nu & = & 0,
\end{eqnarray}
which represent energy conservation and the Euler equation, respectively.
Finally, a perfect fluid does not dissipate energy and hence the
entropy of each fluid element is conserved,
\begin{equation}\label{sdot}
  \dot s=0.
\end{equation}

The system (\ref{Uprime}-\ref{sdot}) fully describes the dynamics of
our background spacetime. In our derivation of the perturbation
equations below, they will further enable us to simplify those
coefficients which depend on the background.

\subsection{High-order perfect fluid perturbations}
\label{sec: perfect_fluid_pert}
Our formulation of the fluid perturbations proceeds in two stages. First,
we expand the perturbations of the energy-momentum tensor in terms of
perturbations $\pert{n}{\rho}$, $\pert{n}{s}$ and $\pert{n}{u}^\mu$
of density, entropy and four-velocity, respectively. Second, we need
to expand the latter as series in appropriate tensor spherical harmonics.

The explicit form of the $n^{\rm th}$ perturbation of the energy-momentum
tensor of the perfect fluid (\ref{fluidemtensor}) is obtained by
using standard combinatorial formulas \cite{xPert}.
The resulting expression is given by
\begin{eqnarray}\label{perturbedemtensor}
  \pert{n}{\Psi}_{\mu\nu}\equiv\Delta^n[t_{\mu\nu}]=\sum_{k=0}^{n}
  \left(
  \begin{array}{c}
   n \\ k
  \end{array}
  \right)
  \Big{\{}\pert{k}{p}\pert{n-k}{h}_{\mu\nu}
  \nonumber\\
  +\sum_{j=k}^{n}
  \left(
  \begin{array}{c}
   n-k \\ j-k
  \end{array}
  \right)
  (\pert{k}{\rho}+\pert{k}{p})\;\pert{j-k}{u}_\mu\pert{n-j}{u}_\nu
  \Big{\}},
\end{eqnarray}
with $\pert{n}{u}_\mu\equiv\Delta^n[u_\mu]$. This can be further
expanded along the same lines, using a high-order generalization
of the chain rule on the equation of state,
\begin{eqnarray}
  \label{eq: dnp}
  \pert{n}{p}&=&\sum \frac{n!}{2!^{k_2}...n!^{k_n}2!^{r_2}...n!^{r_n}
       k_1!...k_n!r_1!...r_n!} \nonumber\\&&\!\!\!\!\!\!\times
  \frac{\partial^{(K+R)}p}{\partial\rho^K\partial s^R}
  \pert{1}{\rho}^{k_1}\dots\pert{n}{\rho}^{k_n}\pert{1}{s}^{r_1}\dots
       \pert{n}{s}^{r_n},
\end{eqnarray}
where the sum is restricted to the following $2n$-tuples,
\begin{displaymath}
\left\{\!(k_1,...,k_n,r_1,...,r_n)\in\mathbb{N}_0^{2n};
\sum_{j=1}^{n}j (k_j+ r_j)=n \!\right\},
\end{displaymath}
and we have defined $K\equiv\sum_{j=1}^n k_j$ and $R\equiv\sum_{j=1}^{n} r_j$.
In practice derivatives of the pressure will
be replaced by the sound speed $c_s$
and thermodynamic factor $C$, defined in Eqs.~(\ref{thermo}).
The expression for the pressure perturbation is completed by the harmonic
decomposition of the density and entropy perturbations
\begin{eqnarray}\label{defomega}
  \pert{n}{\rho}&\equiv& \rho \sum_{l,m}\pert{n}{\omega}_l^m Z_l^m, \\\label{defsigma}
  \pert{n}{s}&\equiv& \sum_{l,m} \pert{n}{\sigma}_l^m Z_l^m.
\end{eqnarray}

Our formulation of the perturbations of the four velocity is guided
by the normalization condition $u_{\mu}u^{\mu}=-1$, which holds at any
perturbative order, so that
\begin{equation}
  \Delta^n[u_\mu g^{\mu\nu} u_\nu]=0.
\end{equation}
By applying the Leibniz rule $n$ times and separating the terms linear
in perturbations of order $n$, we can rewrite this equation as
\begin{eqnarray}\label{constraint}\nonumber
  &&-2 \pert{n}{u}_\mu u^{\mu}
      -u_\mu  u_\nu \Delta^n[g^{\mu\nu}]=
      \\\nonumber
  &&\sum_{k=1}^{n-1}\sum_{i=k}^{n}
      \frac{n!}{k!(i-k)!(n-i)!}\Delta^k[g^{\mu\nu}]
      \pert{i-k}{u}_\mu\pert{n-i}{u}_\nu
      \\
  &&+ \sum_{i=1}^{n-1}
      \left(
      \begin{array}{c}
        n \\ i
      \end{array}
      \right)
      \pert{i}{u}_\mu\pert{n-i}{u}^\mu\equiv \pert{n}{\Upsilon}.
\end{eqnarray}
On the other hand, the $n^{\rm th}$ order perturbation of the inverse
metric is given by \cite{BMM06},
\begin{eqnarray}\label{invmetric}
  \Delta^n[g^{\mu \nu}]&=&
    \sum(-1)^{m}\frac{n!}{k_1!...k_m!}
    \nonumber\\
  &\times & \pert{k_1}{h}^{\mu\alpha}\pert{k_2}{h}_\alpha{}^\beta
    ...\pert{k_{m-1}}{h}_\sigma{}^\rho\pert{k_m}{h}_{\rho}{}^{\nu},
\end{eqnarray}
where this sum extends over all sorted partitions of $n$
such that $k_1+\dots+k_m=n$ with $m \leq n$.
In spherical symmetry the background four-velocity
has no angular components ($u^a=0$), so that the
constraint (\ref{constraint}) only contains
non-angular components of the $n^{\rm th}$
perturbation of the four-velocity.
Since the quantity $\pert{n}{\Upsilon}$ defined in
Eq.~(\ref{constraint}) contains only perturbations up to
order $n-1$, we can make the following ansatz,
\begin{eqnarray}\label{ansatz}
  \pert{n}{u}^A&\equiv&\frac{1}{2}\pert{n}{\Upsilon} u^A
  -\frac{1}{2}\Delta^n[g^{AB}]u_B+\pert{n}{\tilde\gamma} n^A,
\end{eqnarray}
that reproduces the first-order formula used in Ref.~\cite{GuMa00}
and trivially satisfies the constraint (\ref{constraint}).
In this way, the three independent components
of the perturbations of the four-velocity are
encoded in the function $\pert{n}{\tilde\gamma}$,
which will be expanded in a harmonic series with coefficients
denoted by $\pert{n}{\gamma}_l^m$, and the two angular
components $\pert{n}{u}^a$, with harmonic coefficients
$\pert{n}{\alpha}_l^m$ and $\pert{n}{\beta}_l^m$.

The object $\pert{n}{\Upsilon}$, polynomial in lower-order
perturbations and vanishing for $n=1$, is an example of a
high-order perturbation source. These sources, whose
computation is one of the main objectives in this article,
have all a similar internal structure, which we now
illustrate at the second-order perturbative level.

We begin by decomposing the four velocity at first order
(cf.~\cite{GuMa00}),
\begin{eqnarray}
  \pert{1}{u}_\mu dx^\mu &\equiv & \sum_{l,m}
     \left(\pert{1}{\gamma}_l^m{} n_A
     +\frac{1}{2}\pert{1}{\cal{K}}_l^m{}_{AB}u^B\right) Z_l^m dx^A
     \nonumber \\
  &+&\sum_{l,m}\Big( \pert{1}{\alpha}_l^m Z_l^m{}_a
     + \pert{1}{\beta}_l^m X_l^m{}_a \Big) dx^a.
  \label{eq: d1u}
\end{eqnarray}
We recall our discussion below Eq.~(\ref{RW}) and emphasize that
all terms in this equation can be considered as gauge invariant.

At second order, we obtain quadratic first-order terms which arise
from the right-hand side of Eq.~(\ref{ansatz}) and the perturbation
is given by
\begin{eqnarray}\nonumber
  \pert{2}{u}_\mu dx^\mu &\equiv & \sum_{l,m}
     \bigg(\pert{2}{\gamma}_l^m n_A +\frac{1}{2}\pert{2}{\cal{K}}_l^m{}_{AB}u^B
     \\\nonumber
  &&\qquad +\sum_{\bar l, \hat l}
     \U{\epsilon}{\bar l}{\bar m}{\hat l}{\hat m}{l}{m}{}_A\bigg) Z_l^m dx^A
     \\
  &+& \Big(\pert{2}{\alpha}_l^m Z_l^m{}_a
     + \pert{2}{\beta}_l^m X_l^m{}_a \Big) dx^a.
  \label{eq: d2u}
\end{eqnarray}
Here we have defined the alternating sign
\begin{equation}
  \epsilon\equiv(-1)^{\bar l+\hat l-l},
\end{equation}
which is a function of the labels $\bar{l}$, $\hat{l}$, $l$,
even though we ommit these in the notation,
and the sum
\begin{equation}\label{sum}
  \sum_{\bar l, \hat l}\equiv
  \sum_{\hat l=0}^{\infty} \,\,
  \sum_{\bar l=0}^{\infty} \,\,
  \sum_{\hat m=-\hat l}^{\hat l} \,\,
  \sum_{\bar m=-\bar l}^{\bar l}
\;,
\end{equation}
with the restrictions $|\bar l-\hat l| \leq l \leq |\bar l+\hat l|$
and $\bar m + \hat m = m$ for given $l$ and $m$.
The source terms $\onlyU{\epsilon}$ can be classified into two groups;
those of the form $\onlyU{+}$ are formed by $polar\times polar$ or
$axial \times axial$ terms, whereas the terms $\onlyU{-}$
consist of mixed $polar\times axial$ terms.
This structure becomes clear in the explicit form of the sources,
\begin{eqnarray}\nonumber
  \U{+}{\bar l}{\bar m}{\hat l}{\hat m}{l}{m}{}_A &=&
     \E{0}{\hat l}{\hat m}{0}{\bar l}{\bar m}{l}
     \bigg\{u_A\Big[\hat\gamma\bar\gamma-\hat\gamma\bar{\cal K}_{BC}n^Bu^C
     \\\nonumber
  &+&\frac{1}{4}\hat{\cal K}^{BD}\bar{\cal K}_{CD}u_Bu^C
     -\hat{\cal K}^{BC}\bar{\cal K}_{CD}u_Bu^D\Big]
     \\\nonumber
  &-&u_B\hat{\cal K}^{BC}\bar{\cal K}_{AC}\bigg\}
\\\nonumber
    &-&\frac{2}{r^2}\E{1}{\hat l}{\hat m}{-1}{\bar l}{\bar m}{l}\bigg\{
     u_A\Big[\hat\alpha\bar\alpha+\hat\beta\bar\beta+\hat\kappa_Bu^B\bar\kappa_Cu^C
     \\
  &-& 2 \hat\kappa_Bu^B\bar\beta
     - \hat\kappa_B\bar\kappa_Cu^Bu^C\Big]
     \!-\!\hat\kappa_Bu^B\bar\kappa_A\!\bigg\}\!,
     \\
  \U{-}{\bar l}{\bar m}{\hat l}{\hat m}{l}{m}{}_A &=&\frac{4i}{r^2}
     \E{1}{\hat l}{\hat m}{-1}{\bar l}{\bar m}{l} u_A
     \bigg\{\hat\alpha\bar\beta+\bar\alpha\hat\kappa^Bu_B\bigg\}.
\end{eqnarray}
Here, the {\em caret} and {\em bar}
denote first-order harmonic components with
harmonic labels $(\hat l, \hat m)$ and $(\bar l, \bar m)$, respectively.
Even though these individual source terms are not symmetric under the
interchange $(\hat l, \hat m)\leftrightarrow(\bar l, \bar m)$, their
sum of type (\ref{sum}) becomes symmetric in (\ref{eq: d2u}).
In order to save space
we will follow this notation for
the explicit form of source terms quadratic in the first-order
perturbations throughout the remainder of this work.

From now on we will simplify our notation for the first- and
second-order case by dropping the labels $n=1$ and $n=2$. The perturbative
order will become clear from the context in the following equations.
Note also that the expressions for the first-order perturbations
can always be reconstructed from their second-order counterparts
by dropping the quadratic source terms.

With the perturbations of the four-velocity given by
Eqs.~(\ref{eq: d1u}) and (\ref{eq: d2u}), we
can perform the harmonic decomposition of the second-order
perturbation of the energy-momentum tensor (\ref{perturbedemtensor}).
The axial components are given by
\begin{eqnarray}\label{em1}
  \tilde\psi_l^m\!{}_A &=& \beta_l^m (p+\rho) u_A
      \!-\!i \sum_{\bar l, \hat l}
  \Es{-\epsilon}{\bar l}{\bar m}{\hat l}{\hat m}{l}{m}{}_A,
  \\\label{em2}
  \psi_l^m &=& -i \sum_{\bar l, \hat l}
  \Es{-\epsilon}{\bar l}{\bar m}{\hat l}{\hat m}{l}{m},
\end{eqnarray}
where $\cal E$ are first-order quadratic sources that will be given below.
Similarly, we obtain the polar part
\begin{eqnarray}\nonumber
  \Psi_l^m\!{}_{AB} &=& (\rho+p)\bigg[
      \gamma_l^m(u_An_B+n_Au_B) \\\nonumber
  &+&\frac{1}{2}({\cal K}_l^m{}_{AC}u_B+{\cal K}_l^m{}_{BC}u_A)
      \bigg]u^C + p{\cal K}_l^m{}_{AB} \\\nonumber
  &+& \rho\omega_l^m (u_Au_B+c_s^2n_An_B)+C\rho\sigma_l^mn_An_B
     \\\label{em3}
  &+&\sum_{\bar l, \hat l}
     \Es{\epsilon}{\bar l}{\bar m}{\hat l}{\hat m}{l}{m}{}_{AB},
     \\\label{em4}
  \Psi_l^m\!{}_A &=& (p+\rho)\alpha_l^m u_A +
     \sum_{\bar l, \hat l}
     \Es{\epsilon}{\bar l}{\bar m}{\hat l}{\hat m}{l}{m}{}_A,
     \\\nonumber
  {\tilde\Psi}_l^m &=& p {\cal K}_l^m +c_s^2\rho\omega_l^m + C\rho\sigma_l^m
     \\\label{em5}
  &+&\sum_{\bar l, \hat l}
     \Estilde{\epsilon}{\bar l}{\bar m}{\hat l}{\hat m}{l}{m},
     \\\label{em6}
  \Psi_l^m &=& \sum_{\bar l, \hat l}
     \Es{\epsilon}{\bar l}{\bar m}{\hat l}{\hat m}{l}{m}.
\end{eqnarray}
We note that the polarity sign, that appears as a left superindex
of the axial sources is $-\epsilon$, whereas
that of the polar sources is $\epsilon$. Also,
some source terms appear in the same form in the polar and axial equations,
as for example on the right-hand sides of
Eqs.~(\ref{em1}) and (\ref{em4}) or Eqs.~(\ref{em2}) and (\ref{em6}).
This is a general feature of quadratic sources in second-order
perturbation theory; see for example \cite{BMM06}.

We conclude this discussion by giving the sources explicitly
in terms of first-order perturbations and $E$-coefficients,
\begin{widetext}
\begin{eqnarray}\nonumber
  \Es{+}{\bar l}{\bar m}{\hat l}{\hat m}{l}{m}{}_{AB} &=&
     \frac{4}{r^2}(p+\rho)\E{1}{\hat l}{\hat m}{-1}{\bar l}{\bar m}{l}
     \Big{\{}
     \hat\kappa^C\bar\kappa_{(A}u_{B)}u_C
     -u_Au_B\Big[\hat\alpha\bar\alpha+\hat\beta\bar\beta-2\hat\beta\bar\kappa_Cu^C\Big]
     \Big{\}}
     \\\nonumber
  &+& \E{0}{\hat l}{\hat m}{0}{\bar l}{\bar m}{l}
     \bigg{\{}
     2\rho c_s^2\hat{\omega}\bar{\cal K}_{AB}+2\rho\bar{\gamma}
     \left[C \hat{\sigma}
     +\hat{\omega} \left(1+c_s^2\right)\right](n_B u_A+n_A u_B)
     +2 C \rho\hat{\sigma} \Big(\bar{\cal K}_A{}^C n_Bn_C
     +\bar{\cal K}_{C[B} u_{A]}u^C\Big)
     \\\nonumber
  &+&2\rho\hat{\omega}\left(1+c_s^2\right)\bar{\cal K}_{C(B}u_{A)} u^C
     +(p+\rho)\bigg[2 \bar{\gamma} \hat{\gamma } (n_A n_B+u_A u_B)
     -2\hat{\cal K}^{CD}\bar{\cal K}_{D(A}u_{B)} u_C
     \\\nonumber
  &+&2\bar{\gamma} \hat{\cal K}_{C(A}n_{B)} u^C-2\bar{\gamma}
     \hat{\cal K}_{CD}n^Cu^Du_A u_B
     +\frac{1}{2} \hat{\cal K}_{AC} u^C\bar{\cal K}_{BD} u^D
     -\frac{3}{2}\hat{\cal K}_C{}^F\bar{\cal K}_{DF} u^C u^D u_A u_B\bigg]
     \\\label{mattersource1}
  &+&\rho n_A n_B\left(\bar{\sigma}\hat{\sigma}\frac{\partial C}{\partial s}
     +2\bar{\omega} \hat{\sigma}\frac{\partial c_s^2}{\partial s}
     +\rho\bar{\omega}\hat{\omega}\frac{\partial c_s^2}{\partial\rho}\right)
     \bigg{\}},
     \\
  \Es{-}{\bar l}{\bar m}{\hat l}{\hat m}{l}{m}{}_{AB} &=&
     \frac{8 i}{r^2}\E{1}{\hat l}{\hat m}{-1}{\bar l}{\bar m}{l}
     \bar\alpha(\hat\kappa^Cu_C-\hat\beta)(p+\rho)u_Au_B,
     \\
  \Es{+}{\bar l}{\bar m}{\hat l}{\hat m}{l}{m}{}_{A} &=&
     2 \E{0}{\hat l}{\hat m}{1}{\bar l}{\bar m}{l}
     \bar\alpha
     \bigg{\{}
     (p+\rho)\left[
     \hat\gamma n_A + \frac{1}{2}\hat{\cal K}_{AB}u^B
     \right]
     +\rho\hat\omega (1+c_s^2)u_A + C\rho\hat\sigma u_A
     \bigg{\}},
     \\
  \Es{-}{\bar l}{\bar m}{\hat l}{\hat m}{l}{m}{}_{A} &=&
     2 i \E{0}{\hat l}{\hat m}{1}{\bar l}{\bar m}{l}
     \bigg{\{}
     (p+\rho)\left[\hat\gamma n_A+\frac{1}{2}\hat{\cal K}_{AB}u^B\right]\bar\beta
     +\rho\bar\beta [(c_s^2+1)\hat\omega+C\hat\sigma] u_A
     +\rho (C\hat\sigma+c_s^2\hat\omega)\bar\kappa_A
     \bigg{\}},
     \\
  \Estilde{+}{\bar l}{\bar m}{\hat l}{\hat m}{l}{m} &=&
     -\frac{2}{r^2}(p+\rho)\E{1}{\hat l}{\hat m}{-1}{\bar l}{\bar m}{l}
     \Big{\{}
     \hat\alpha\bar\alpha + \hat\beta\bar\beta
     \Big{\}}
     +2 \rho \E{0}{\hat l}{\hat m}{0}{\bar l}{\bar m}{l}
     \bigg{\{}
     \hat{\cal K}(\bar\omega c_s^2 + \bar\sigma C)
     +\hat\omega\bar\sigma\frac{\partial c_s^2}{\partial s}
     +\frac{1}{2}\hat\sigma\bar\sigma\frac{\partial C}{\partial s}
     +\frac{\rho}{2}\hat\omega\bar\omega\frac{\partial c_s^2}{\partial \rho}
     \bigg{\}},
     \\
  \Estilde{-}{\bar l}{\bar m}{\hat l}{\hat m}{l}{m} &=&
     \frac{4i}{r^2}\E{1}{\hat l}{\hat m}{-1}{\bar l}{\bar m}{l}
     (p+\rho)\hat\alpha\bar\beta,
     \\
  \Es{+}{\bar l}{\bar m}{\hat l}{\hat m}{l}{m} &=&
     2 \E{1}{\hat l}{\hat m}{1}{\bar l}{\bar m}{l}
     (\hat\alpha\bar\alpha-\hat\beta\bar\beta)(p+\rho),
     \\\label{mattersource8}
  \Es{-}{\bar l}{\bar m}{\hat l}{\hat m}{l}{m} &=&
     4 i \E{1}{\hat l}{\hat m}{1}{\bar l}{\bar m}{l}
     \hat\alpha\bar\beta(p+\rho).
\end{eqnarray}
\end{widetext}

\section{Second-order evolution equations}
\label{sec: evol_eqs}

We now turn our attention to the evolution equations.
The results derived in the previous sections facilitate the formulation
of the evolution equations at arbitrary order. In order to keep the complexity
of the equations in this article at a manageable level, however,
we will restrict this discussion to the simplest nonlinear case,
the evolution equations for second-order perturbations.

Note again that our calculation represents a generalization
of the first-order results of Ref.~\cite{GuMa00}.
Because the equations at second order share
the linear part with those at first order, our main task will be
the computation of the
additional source terms quadratic in the first-order perturbations.

In the Einstein equations
these quadratic source terms contain two parts; (i) terms arising
from the second-order perturbation of the Einstein tensor
which were denoted by
$\Source{\epsilon}{\bar{l}}{\bar{m}}{\hat{l}}{\hat{m}}{l}{m}$
in their original derivation in Ref.~\cite{BMM06};
(ii) sources
$\Es{\epsilon}{\bar{l}}{\bar{m}}{\hat{l}}{\hat{m}}{l}{m}$,
obtained
from the perturbation of the energy-momentum tensor,
which have been given explicitly in
Eqs.~(\ref{mattersource1}-\ref{mattersource8}) for the case of
perfect fluid matter. It is convenient to combine those two types as
\begin{equation}
  {{}^{\mbox{\tiny $({\epsilon})\!$}}{\cal P}_{\bar{l}}^{\bar{m}}
      {}_{\hat{l}}^{\hat{m}}{}_{l}^{m}}\equiv
      8 \pi \Es{\epsilon}{\bar{l}}{\bar{m}}{\hat{l}}{\hat{m}}{l}{m}
      -\Source{\epsilon}{\bar{l}}{\bar{m}}{\hat{l}}{\hat{m}}{l}{m},
\end{equation}
for all tensorial, vectorial and scalar sources.
The perturbative evolution equations for the matter can be computed
straightforwardly
by direct perturbation of their nonlinear counterparts. For simple
systems, including perfect fluids, they can also be
obtained from the equations of
energy-momentum conservation. In particular, the quadratic sources
for the fluid equations of motion can be computed from the
${\cal I}_{\bar{l}}^{\bar{m}}{}_{\hat{l}}^{\hat{m}}{}_{l}^{m}$
sources of energy-momentum conservation defined in
Eqs.\ (100-102) of Ref. \cite{BMM06}.

Once again, in order to simplify our notation, we will omit from now on
the harmonic labels $\{l,m,\hat l, \hat m, \bar l, \bar m\}$ from the
source names and merely write
${}^{\mbox{\tiny $({\epsilon})\!$}}{\cal P}$,
${}^{\mbox{\tiny $({\epsilon})\!$}}{\cal P}_A$,
${}^{\mbox{\tiny $({\epsilon})\!$}}{\cal P}_{AB}$, and so on.
We also omit the labels $\{l,m\}$ from the second-order perturbations.
They will be distinguished from those at first order because the latter
are always denoted with hat or bar overscripts.

Let us summarize the variables that will be used
to describe the different perturbative degrees of freedom.
The fundamental variables describing density $\omega$, entropy
$\sigma$ and velocity perturbations $\{\alpha, \beta, \gamma\}$,
have already been defined in Eqs. (\ref{defomega}), (\ref{defsigma})
and (\ref{eq: d2u}) respectively. All other tensorial
perturbative variables will be decomposed in the natural background
frame provided by the fluid velocity $(u^A, n^A)$ as follows.
We encode the axial metric information in the scalars
$\{\delta, \lambda\}$, defined as
\begin{equation}\label{framevector}
\kappa_A\equiv\delta u_A +\lambda n_A,
\end{equation}
whereas the polar tensor ${\cal K}_{AB}$ is decomposed in three
scalar functions,
\begin{eqnarray}\nonumber
{\cal K}_{AB} & \equiv & \eta (n_An_B-u_Au_B) + \phi (u_Au_B+n_An_B)
\\&+&\psi (u_An_B+n_Au_B).
\label{dectensor}
\end{eqnarray}
Note that the scalar $\psi$ introduced here, and that will be used
in the rest of the article, is different from the harmonic function
$\psi$ in the decomposition of the perturbations of the
energy-momentum tensor (\ref{emdecomposition}). In addition,
for future convenience, the metric component function $\phi$ will
be replaced by the new variable $\chi$, defined by
\begin{equation}\label{defchi}
\phi\equiv \chi+{\cal K}-\eta.
\end{equation}
Finally, the fourth polar geometric degree of freedom will be
described by the scalar gauge-invariant ${\cal K}$.

Regarding the linear evolution equations, we sill separate the
polar and axial parts for arbitrary (second-order) harmonic label $l$.
As shown in Ref. \cite{GuMa00},
for those cases with $l\geq 2$, the axial sector reduces to two independent
equations, one for the axial gravitational wave ($\Pi$, defined below)
and the other for a fluid
perturbation ($\beta$) describing rotation. On the other hand, the polar
sector, also for $l\geq 2$, contains four fundamental equations for four
variables, respectively describing the polar gravitational wave ($\chi$),
the sound wave (${\cal K}$), the entropy perturbation ($\sigma$) and
non-rotational tangential fluid motion ($\psi$). All other metric and
fluid perturbations can be reconstructed from those using the constraint
equations. In particular, the function $\eta$ turns out to be vanishing
in the case $l\geq 2$. The particular cases $l=0, 1$ are
non-radiative and the corresponding equations present a different
structure. In particular it is not possible to construct
gauge invariants for $l=0,1$, and we will need to resort to adequate
gauge fixing \cite{ThCa67, CaTh70}.

\subsection{Axial perturbations}

\subsubsection{The case $l\geq 2$}

There is a single axial fluid perturbation $\beta$
(defined in Eq.~(\ref{eq: d2u})), which obeys
the following transport equation,
equivalent to the axial part of the perturbed equation of energy-momentum
conservation (cf.~Eq.~(99) of Ref.~\cite{BMM06}),
\begin{eqnarray}
 & & \dot\beta-c_s^2(\mu+2U)\beta= i\;\sum_{\bar l, \hat l} {}^{(-\epsilon)}{\cal B} \equiv
      \nonumber\\
 & &  \qquad \equiv \frac{i}{p+\rho}\sum_{\bar l,\hat l}
      \Bigg{\{}{}^{(-\epsilon)}{\cal I}
      -\frac{(l-1)(l+2)}{2r^2}{}^{(-\epsilon)}{\cal E}
      \nonumber \\ \label{matteraxial} 
 & &  \qquad \qquad \qquad \qquad +\frac{1}{r^2}\Big[r^2{}^{(-\epsilon)}{\cal E}_A \Big]^{|A}\Bigg{\}}.
\end{eqnarray}
The source term ${}^{(\epsilon)}{\cal B}$ is given
in expanded form as follows:

\begin{widetext}
\begin{eqnarray}
{}^{(-)}{\cal B}
&=& i\frac{2(l-1)(l+2)}{r^2} \bar\alpha \hat\beta
\E{1}{\hat l}{\hat m}{1}{\bar l}{\bar m}{l}
 \nonumber \\
&-&i\,
\E{0}{\hat l}{\hat m}{1}{\bar l}{\bar m}{l}
\biggl\{
  \bar\beta' \bigl(2 \hat\gamma + \hat\psi\bigr)
+ \bar\beta
\biggl(- \frac{1}{c_s^2}
\Bigl(\frac{C s' \rho}{\rho+p} + \nu \Bigr) (2 \hat\gamma + \hat\psi)
+ \frac{2 \rho}{\rho+p} (\dot{\hat\omega} +c_s^2\dot{\hat\omega}+ C \dot{\hat\sigma} )
+ 2\hat\gamma'+\hat\psi'
+ \dot{\hat\chi}+3\dot{\hat{\mathcal{K}}}
\nonumber \\
&+& 2 \frac{1+c_s^2}{\rho+p}(\mu+2U)
 (\rho C \hat\sigma + \rho c_s^2 \hat\omega - p \hat\omega)
+2 W (2\hat\gamma+\hat\psi)
- 2 (\mu+2U) (\hat\sigma \frac{\partial c_{s}^{2}}{\partial s}+ \hat\omega \rho\frac{\partial c_{s}^{2}}{\partial \rho})
\biggr)
\biggr\},
\label{Bminus}
\end{eqnarray}
\begin{eqnarray}
{}^{(+)}{\cal B}
&=& \frac{(l-1)(l+2)}{r^2}
\E{1}{\hat l}{\hat m}{1}{\bar l}{\bar m}{l}
(\hat\alpha \bar\alpha -  \hat\beta \bar\beta)
\nonumber \\
&+&\E{0}{\hat l}{\hat m}{1}{\bar l}{\bar m}{l}
\biggl\{\frac{\bar{l} (\bar{l}+1)}{r^2} 2\bar\beta \hat\delta
 + 2 \bar\gamma \hat\psi
 - \frac{2 \rho}{\rho+p} \bar{\mathcal{K}} (c_{s}^{2} \hat\omega + C \hat\sigma)
 + \frac{\rho}{\rho+p} (2\hat\eta-\hat\chi) \bar\omega (1+c_s^2)
-\bar\alpha' (2 \hat\gamma + \hat\psi)
\nonumber \\
 &-& \bar\alpha (2\hat\gamma' + \hat\psi')
 + \dot{\hat\alpha} \Bigl(
  (\bar\chi - 2 \bar\eta + \bar{\mathcal{K}})
  + \frac{2 \rho}{\rho+p} (\bar\omega + c_{s}^{2} \bar\omega + C \bar\sigma)\Bigr)
+\frac{1}{c_s^2} \bar\alpha \Bigl(\frac{C s' \rho}{\rho+p} + \nu \Bigr) (2 \hat\gamma + \hat\psi)
\nonumber \\
&+& c_s^2 \bar\alpha \bigl(\mu  + 2 U\bigr) (\hat{\mathcal{K}}+\hat\chi-2\hat\eta) 
 - 2 \bar\alpha \dot{\hat\omega} \frac{\rho}{\rho+p} (1+c_s^2)
 - 2 \bar\alpha C \dot{\hat\sigma} \frac{\rho}{\rho+p}
 + 2 \bar\alpha (\mu+2U)  \hat\omega \frac{p}{\rho+p} (1+c_s^2)
\nonumber\\
&+& \bar{\mathcal{K}} \hat\omega \frac{\rho(1+3c_s^2)}{\rho+p}
 - \bar\alpha (\dot{\hat\chi}+3\dot{\hat{\mathcal{K}}})
 - 2 \bar\alpha (2\hat\gamma+\hat\psi) W
 + \frac{\rho C \hat\sigma}{\rho+p} (
   \bar\chi -2\bar\eta + 3\bar{\mathcal{K}}
)
 - 2 \bar\alpha (\mu+2U) \frac{\rho C\hat\sigma}{\rho+p}
\nonumber \\
&+& 2 \bar\alpha (\mu+2U) (\hat\sigma \frac{\partial c_{s}^{2}}{\partial s} + \hat\omega \rho \frac{\partial c_{s}^{2}}{\partial \rho} )
\biggr\} .
\label{Bplus}
\end{eqnarray}
\end{widetext}
In order to simplify this expression we have assumed that the first-order
harmonic numbers $\hat l$ and  $\bar l$ are greater or equal to 2 and,
thus, both $\hat\eta$ and $\bar\eta$ are vanishing, as will be made
explicit after Eq. (\ref{fluideq7bis}). 
We could expand the time derivatives of the first-order perturbations
$\alpha$, $\omega$ and $\sigma$, using their evolution equations,
but it does not lead to any simplification.

At first order the axial perturbation $\beta$ evolves freely on the
background, in the sense that it obeys a transport equation in which
the axial gravitational wave does not appear. We see in (\ref{Bminus})
that the source terms ${}^{(-)}{\cal B}$
extend this fact to second order because they do not contain any axial
metric perturbations, due to a geometrical cancellation of the
$E$-coefficients of $\bar\alpha \hat\delta$.
Such a cancellation does not occur for the ${}^{(+)}{\cal B}$
source terms in (\ref{Bplus}), which contains a term
\begin{equation}
E{}^{1}_{0}{}^{\bar{l}}_{\hat{l}}{}^{\bar{m}}_{\hat{m}}{}_{l}
\frac{\bar{l} (\bar{l}+1)}{r^2} 2\bar\beta \hat\delta .
\end{equation}
This allows a coupling between the axial gravitational wave $\hat\delta$
and a rotational mode $\bar\beta$, generating a second-order rotational
mode $\beta$ for odd $\hat{l}+\bar{l}-l$. Note
that in such a case we have
$E{}^{1}_{0}{}^{\bar{l}}_{\hat{l}}{}^{\bar{m}}_{\hat{m}}{}_{l}
= - E{}^{0}_{1}{}^{\bar{l}}_{\hat{l}}{}^{\bar{m}}_{\hat{m}}{}_{l}$
(see \cite{BMM06}).

At any given perturbative order $n$
the axial gravitational wave is fully described by
the Gerlach-Sengupta master scalar
\begin{equation}\label{Pidefinition}
  \pert{n}{\Pi}\equiv\epsilon^{AB}\left(
      \frac{\pert{n}{\kappa}_A}{r^2}\right)_{|B}.
\end{equation}
For $l\ge 2$ one of the perturbed Einstein equations provides a simple
way to reconstruct the vector $\kappa_A$ from the master scalar $\Pi$:
\begin{eqnarray}\nonumber
  && (l-1)(l+2)\kappa_A=  \\
  && \qquad = 16 \pi r^2 \tilde\psi_A-\epsilon_{AB}(r^4\Pi)^{|B}
  +2 i r^2\sum_{\bar l, \hat l}{}^{(-\epsilon)}S_A , \qquad
\label{vector}
\end{eqnarray}
(cf.\ Eqs.\ (80) and (D5) of \cite{BMM06}),
and hence there is a one-to-one correspondence between $\Pi$ and
$\kappa_A$, assuming the matter perturbations and sources are known.
Expanding this relation in the natural frame, as done in
(\ref{framevector}), results in the following expressions for the
component functions
\begin{eqnarray}\nonumber
  \delta &=& \frac{r^2}{(l-1)(l+2)}\Big(r^2\Pi'+4r^2W\Pi+16\pi\beta(p+\rho)
      \\\label{reconstruct1}
  && \qquad +2i \sum_{\bar l, \hat l}{}^{(-\epsilon)}{\cal P}_Au^A\Big),
        \\\nonumber
  \lambda &=& -\frac{r^2}{(l-1)(l+2)}\Big(r^2\dot{\Pi}+4r^2 U \Pi
      \\\label{reconstruct2}
  &&+ \quad 2 i\sum_{\bar l, \hat l}{}^{(-\epsilon)}{\cal P}_An^A\Big).
\end{eqnarray}

One further differentiation of (\ref{vector}) gives 
an evolution equation for the
axial sector of the Einstein equations, which at second order is
\begin{eqnarray}\nonumber
  -\left[\frac{1}{2r^2}(r^4{\Pi})^{|A}\right]_{|A}
      +\frac{(l-1)(l+2)}{2}{\Pi}
      = \\ =
  8\pi\epsilon^{AB}{\tilde\psi}_{A|B}
      +i \epsilon^{AB}\sum_{\bar{l},\hat{l}}
      {}^{(-\epsilon)}S_{A|B}.
  \label{eq: pert_metr_v1}
\end{eqnarray}
By expanding the energy-momentum perturbations in terms of the
fundamental fluid variables (\ref{em1}), this equation becomes
\begin{eqnarray}\nonumber
  -\left[\frac{1}{2r^2}(r^4\Pi)^{|A}\right]_{|A}
       +\frac{(l-1)(l+2)}{2}\Pi
       = \\ \nonumber \quad =
  8\pi(p+\rho)\beta^{\prime}
       -\frac{8\pi}{c_s^2}[\rho C s'+\nu(p+\rho)]\beta
       \\\label{master}
  -i \epsilon^{AB}
       \sum_{\bar{l},\hat{l}}
       {}^{(-\epsilon)}{\cal P}_{A|B}.
\end{eqnarray}
The expanded form of this source is too large to be given here.

In summary, the evolution of the second-order axial perturbations
for $l\ge 2$ is given by two coupled equations; the matter equation
(\ref{matteraxial}) determines the evolution of $\beta$
and the master equation (\ref{master}) that of the Gerlach and
Sengupta scalar $\Pi$.
Reconstruction of the metric is facilitated by using
Eqs.~(\ref{reconstruct1}-\ref{reconstruct2}).

\subsubsection{The case $l=1$}\label{axial:l=1}

For $l=1$ the right-hand side of the matter equation (\ref{matteraxial})
slightly simplifies but the overall structure remains that of a transport
equation for $\beta$,
\begin{equation}\label{betamatter}
  (p+\rho)[\dot\beta-c_s^2(\mu+2U)\beta]=i\sum_{\bar l,\hat l}
      \Big{\{}{}^{(-\epsilon)}{\cal I}
      +\frac{1}{r^2}\big(r^2{}^{(-\epsilon)}{\cal E}_A \big)^{|A}\Big{\}},
\end{equation}
whose sources are given by (\ref{Bminus}) and (\ref{Bplus}) particularizing
to $l=1$. However, the treatment of the gravitational part is rather
different.
The metric perturbation equation (\ref{vector}) is still valid for $l=1$
and now becomes an equation for $\Pi$,
\begin{equation}
  \frac{1}{2r^2}\epsilon_{AB}(r^4\Pi)^{|B}=
      8 \pi\tilde\psi_A+i\sum_{\bar l, \hat l}{}^{(-\epsilon)}S_A .
\end{equation}
This allows direct integration of $\Pi$ from the knowledge of $\beta$ and
the sources, reflecting the non-existence of dipole gravitational waves.
This is clearer by projecting the equation onto the frame vectors $u^A$
and $n^A$ and expanding the fluid term according to Eq.~(\ref{em1}):
\begin{eqnarray}\label{betaalge}
  \frac{r^2}{2}(\Pi'+4W\Pi)&=&-8\pi(p+\rho)\beta\!-\!i\sum_{\bar l, \hat l}
      {}^{(-\epsilon)}{\cal P}_Au^A\!,\\
  \frac{r^2}{2}(\dot\Pi+4U\Pi)&=&-i\sum_{\bar l, \hat l}
      {}^{(-\epsilon)}{\cal P}_An^A.
      \label{eq: Pidot}
\end{eqnarray}
In practice we plan to first integrate $\beta$ using the evolution equation
(\ref{betamatter}) and then obtain $\Pi$ from the constraint (\ref{betaalge}).
The former implies dividing by the factor $(p+\rho)$ which, at least
for some equations of state, might vanish at the surface of a fluid ball but
the solution can be shown to be nonsingular by power series expansions
around the surface.

A further difference is that now we cannot reconstruct $\kappa_A$ from
$\Pi$ using the perturbed Einstein equations. We need to integrate the
definition (\ref{Pidefinition}) of $\Pi$.
We proceed as follows. First, we note that
any two-dimensional vector can be expressed in terms of two scalar functions
$y(x^A)$ and $z(x^A)$ by writing
\begin{equation} \label{kappafg}
  \frac{1}{r^2}\kappa_A\equiv y_{|A}+\epsilon_A{}^Bz_{|B}=
      -(\dot y +z')u_A + (y'+\dot z)n_A.
\end{equation}
We insert this expression into the
definition of the master scalar (\ref{Pidefinition}) and obtain
\begin{equation}\label{forg}
  \Pi=-z^{|A}{}_{|A}=\ddot{z}+\mu \dot z-\nu z'-z''.
\end{equation}
Finally, we note that in the special case $l=1$, the symmetric trace free
tensor $X_l^m{}_{ab}$ defined in Eq.~(\ref{Xlm}) vanishes, so that the
condition $\pert{n}{h}_l^m=0$ in Eq.~(\ref{RW}) no longer provides
a gauge condition. In RW gauge for $l=1$ there thus remains one axial
degree of freedom. The function $y$ represents this freedom and a gauge
transformation $y \rightarrow y+\xi$ causes a change in the vector
$\kappa_A$
\begin{equation}
  \kappa_A\longrightarrow\kappa_A + r^2 \xi_{|A} + \dots.
\end{equation}
Here the dots indicate source terms quadratic in first-order perturbations
which are given explicitly in Ref.~\cite{BMM07}. The resulting change
in the component functions $\delta$ and $\lambda$
defined in Eqs.~(\ref{reconstruct1}) and (\ref{reconstruct2}) is
\begin{eqnarray}
  \delta\longrightarrow\delta - r^2 \dot\xi +\dots,\\
  \lambda\longrightarrow\lambda + r^2 \xi' +\dots.
\end{eqnarray}
We are therefore able to fix the gauge by eliminating either
$\delta$ or $\lambda$ or any
combination of the two. It turns out to be most convenient to
set $\lambda=0$. Equation (\ref{kappafg}) then gives $y'=-\dot{z}$,
which allows us to integrate $y$ from $z$ and the values of
$y$ at $r=0$ (a residual gauge corresponding to a generator
$\xi$ which obeys $\xi'=0$). In this gauge, the vector $\kappa_A$
is given by,
\begin{equation}
  \kappa_A=-r^2(\dot y + z')u_A,
\end{equation}
where $z$ is obtained from equation (\ref{forg}) and $y'=-\dot z$.

\subsection{Polar perturbations}

Following decomposition (\ref{dectensor}), we can schematically write the
second-order Einstein equations as follows.
We indicate in each case the range of values of $l$, the label
of the second-order perturbation, for which the various equations hold.
For $l\geq 0$,
\begin{eqnarray}
  u^An^B E_{AB}[{\cal K}] &=& -8\pi (\rho + p)\gamma +4\pi (\rho - p)\psi
      \nonumber\\\label{fluideq1}
  &+&\sum_{\bar{l},\hat{l}}
      {}^{(\epsilon)}{\cal P}{}_{AB}u^An^B,
      \\
  u^Au^B E_{AB}[{\cal K}] &=& 8\pi\rho\omega + 8\pi\rho (\eta-\phi)
      \nonumber\\\label{fluideq2}
  &+&\sum_{\bar{l},\hat{l}}
      {}^{(\epsilon)}{\cal P}{}_{AB}u^Au^B,
      \\
  n^An^B E_{AB}[{\cal K}] &=& 8\pi\rho (c_s^2\omega+C\sigma)+8\pi p(\eta+\phi)
      \nonumber\\\label{fluideq3}
  &+&\sum_{\bar{l},\hat{l}}
      {}^{(\epsilon)}{\cal P}{}_{AB}n^An^B,
      \\
  \tilde E[{\cal K}] &=& 8\pi\rho (c_s^2\omega+C\sigma)+8\pi p{\cal K}
      \nonumber\\\label{fluideq4}
  &+&\sum_{\bar{l},\hat{l}}
      {}^{(\epsilon)}\tilde{\cal P}.
\end{eqnarray}
For $l\geq 1$,
\begin{eqnarray}\label{fluideq5}
  u^A E_{A}[{\cal K}] &=& -(\rho+p)\alpha
      +\sum_{\bar{l},\hat{l}}
      {}^{(\epsilon)}{\cal P}{}_{A}u^A,
      \\\label{fluideq6}
  n^A E_{A}[{\cal K}] &=&\sum_{\bar{l},\hat{l}}
      {}^{(\epsilon)}{\cal P}{}_{A}n^A.
\end{eqnarray}
And finally, for $l\geq 2$,
\begin{equation}\label{fluideq7}
  E[{\cal K}] =\sum_{\bar{l},\hat{l}}
      {}^{(\epsilon)}{\cal P}.
\end{equation}
The $E$ are the Gerlach and Sengupta linear differential operators which
act on the polar metric perturbations $\{{\cal K}_{AB}, {\cal K}\}$ and
are defined explicitly in Eqs. (D1-D4) of Ref.~\cite{BMM06}.

\subsubsection{The case $l\geq 2$}

By expanding the part of equation (\ref{fluideq7})
which is linear in the second-order perturbations we arrive at the
simple expression
\begin{equation}\label{fluideq7bis}
  \eta =-\sum_{\bar{l},\hat{l}} {}^{(\epsilon)}{\cal P}.
\end{equation}
The scalar $\eta$ is thus given directly in terms of the first-order
perturbations. Note that, as commented at the beginning of this section,
in the case of first-order perturbation theory there are no source terms
and the first-order $\eta$ vanishes. Even though, Eq. (\ref{fluideq7bis})
does not exist for $l=0,1$ and hence, in these cases, the first-order
$\eta$ is nonvanishing.

Motivated by the linearized equations of energy-momentum conservation,
we can take linear combinations of the system
(\ref{fluideq1}-\ref{fluideq6}) such that it can be rewritten as
\begin{eqnarray}\nonumber
  l\geq 0:\\\label{polareq1}
  8\pi(p+\rho)\gamma = (\dot{\cal K})'+C_{\gamma}
      +\sum_{\bar l, \hat l} {}^{(\epsilon)}C_{\gamma},\\\label{polareq2}
  8\pi\rho\omega = -{\cal K}''+2U\psi'+C_{\omega}
      +\sum_{\bar l, \hat l} {}^{(\epsilon)}C_{\omega},\\\nonumber
  l\geq 1:\\\label{polareq3}
  16\pi(p+\rho)\alpha = \psi'+C_{\alpha}
      +\sum_{\bar l, \hat l} {}^{(\epsilon)}C_{\alpha},\\\label{polareq4}
  -\ddot\chi+\chi''+2(\mu-U)\psi'=S_{\chi}
      +\sum_{\bar l, \hat l} {}^{(\epsilon)}S_{\chi},\\\label{polareq5}
  -\ddot{\cal K}+c_s^2{\cal K}''-2c_s^2U\psi'=S_{\cal K}
      +\sum_{\bar l, \hat l} {}^{(\epsilon)}S_{\cal K},\\\label{polareq6}
  -\dot\psi = S_{\psi}+\sum_{\bar l, \hat l} {}^{(\epsilon)}S_{\psi},
\end{eqnarray}
where $C_\gamma$, $C_\omega$, $C_\alpha$, $S_\chi$, $S_{\cal K}$ and
$S_\psi$ are linear functions of the quantities $\chi$, ${\cal K}$ and
$\eta$, their derivatives $\{\dot{\chi}, \chi',\dot{\cal K}, {\cal K}',
\dot{\eta}, \eta', \eta''\}$ and of $\psi$ and $\sigma$.
These sources are identical to those given
in Ref.~\cite{GuMa00} and are included in App.~\ref{fluidsources}
for completeness.
We have also introduced the following linear combinations of quadratic
sources,
\begin{eqnarray}
  {}^{(\epsilon)}C_\gamma& \equiv &{}^{(\epsilon)}{\cal P}_{AB}u^An^B,\\
  {}^{(\epsilon)}C_\omega& \equiv & - {}^{(\epsilon)}{\cal P}_{AB}u^Au^B,\\
  {}^{(\epsilon)}C_\alpha& \equiv & 2{}^{(\epsilon)}{\cal P}_Au^A,\\
  {}^{(\epsilon)}S_\chi& \equiv &2{}^{(\epsilon)}\tilde{\cal P}
      +4\Big[{}^{(\epsilon)}{\cal P}_An^A\Big]'-2{}^{(\epsilon)}{\cal P}_{AB}n^An^B
      \nonumber\\
  &+&4(2\nu-W){}^{(\epsilon)}{\cal P}_An^A,\\
      {}^{(\epsilon)}S_{\cal K}& \equiv &
      \Big(-c_s^2u^Au^B+n^An^B\Big){}^{(\epsilon)}{\cal P}_{AB}
\nonumber\\&+&4W{}^{(\epsilon)}{\cal P}_An^A,\\
  {}^{(\epsilon)}S_\psi & \equiv &-2{}^{(\epsilon)}{\cal P}_An^A.
\end{eqnarray}
In order to illustrate the kind of quadratic terms that appear
in the above equations, here we also present
explicitly those corresponding to the last equation
for the variable $\psi$,
\begin{widetext}
\begin{eqnarray}
{}^{(+)}S_\psi&=&
\frac{2}{r^2}
\E{2}{\hat l}{\hat m}{-1}{\bar l}{\bar m}{l}
\Bigl\{- \bar\delta \dot{\hat\lambda} + \hat\delta \dot{\bar\lambda} + 
\mu\bar\delta \hat\lambda  - \mu\hat\delta \bar\lambda  + \bar\delta \hat\delta'
+ \hat\delta \bar\delta' - 2 \bar\lambda \hat\lambda' - 2 W\hat\delta \bar\delta
+ 2 W\hat\lambda \bar\lambda \Bigr\}
\nonumber\\
&+&
\E{1}{\hat l}{\hat m}{0}{\bar l}{\bar m}{l}
\biggl\{2 \bar{\mathcal{K}} \hat{\mathcal{K}}' + 2 \hat{\mathcal{K}} \bar{\mathcal{K}}'
 + 2 \hat\psi \bigl(\dot{\bar\chi} + \dot{\bar{\mathcal{K}}}\bigr) - 2 \bigl(\dot{\bar\psi} + 2 \mu  \bar\psi\bigr) \bigl(\hat\chi + \hat{\mathcal{K}}\bigr) - 16 \pi \hat\alpha \bigl(p + \rho\bigr) \bigl(2\bar\gamma - \bar\psi\bigr) \nonumber \\
 &+& \frac{\bar{l}\bigl(1 + \bar{l}\bigr)}{r^{2}}
\Bigl[3 \hat\delta \dot{\bar\lambda}- \bar\delta \dot{\hat\lambda}  + 2 \dot{\hat\delta} \bar\lambda + \mu\bar\delta \hat\lambda   + 3 \mu\hat\delta \bar\lambda   + \nu\hat\lambda \bar\lambda
+ \bar\delta \hat\delta' + 3 \hat\delta \bar\delta' - 4 U\hat\delta \bar\lambda
- 2 (3 W-\nu ) \hat\delta \bar\delta +  (2 W+\nu)\hat\lambda \bar\lambda\Bigr]
\nonumber\\
 &+& 2 \bigl(\bar\chi + \bar{\mathcal{K}}\bigr)\bigl(\dot{\hat\psi} + \hat\chi' + \hat{\mathcal{K}}' + 2 \mu  \hat\psi - 2 W\hat\chi - 2 W\hat{\mathcal{K}}\bigr)  - 2 \bar\psi \bigl(\dot{\hat\chi} + \dot{\hat{\mathcal{K}}} + \hat\psi' - 2 W\hat\psi \bigr)\biggr\},
\end{eqnarray}
\begin{eqnarray}
{}^{(-)}S_\psi&=&
-\frac{2i}{r^2}
\E{-1}{\hat l}{\hat m}{2}{\bar l}{\bar m}{l}
\biggl\{\bar\chi \hat\lambda + \bar{\mathcal{K}} \hat\lambda + \hat\chi \bar\lambda + \hat{\mathcal{K}} \bar\lambda -  \bar\delta \hat\psi -  \hat\delta \bar\psi\biggr\} \nonumber \\
 &-& 2i
\E{1}{\hat l}{\hat m}{0}{\bar l}{\bar m}{l}
\biggl\{\hat\lambda \Bigl(\ddot{\bar\chi} + 3 \ddot{\bar{\mathcal{K}}}\Bigr) -  \ddot{\hat\lambda} + \dot{\hat\delta}' \bigl(\bar\chi + \bar{\mathcal{K}}\bigr)
 + \hat\delta \dot{\bar{\mathcal{K}}}' + \hat\lambda \Bigl(2 \dot{\bar\psi}' + \bar\chi''\Bigr)
+  \mu '\hat\lambda \bar\psi -  \bar\psi \Bigl(\dot{\hat\lambda}' + \hat\delta''\Bigr)
+ \nu ' \Bigl[3 \hat\lambda \bigl(\bar\chi + \bar{\mathcal{K}}\bigr) -  \hat\delta \bar\psi\Bigr]
\nonumber \\  
&-&  \dot{\nu } \Bigl[\hat\delta \bigl(\bar\chi + \bar{\mathcal{K}}\bigr) - 2 \hat\lambda \bar\psi\Bigr]
+ \frac{2M}{r^{3}} \hat\delta \bar\psi
+ \frac{\hat{l}\bigl(1 + \hat{l}\bigr)}{r^{2}} \bar{\mathcal{K}} \hat\lambda
+ \frac{\bar{l}\bigl(1 + \bar{l}\bigr)}{2 r^{2}}
\Bigl(\bar{\mathcal{K}} \hat\lambda- \bar\chi \hat\lambda + \hat\chi \bar\lambda + 3 \hat{\mathcal{K}} \bar\lambda -  \bar\delta \hat\psi + \hat\delta \bar\psi\Bigr)
- 8 \pi \rho \,\hat\delta \bar\psi
\nonumber\\
&+& 8 \pi \bigl(p + \rho\bigr) \bigl(2 \bar\gamma -  \bar\psi\bigr)\hat\beta 
- 4 \pi \hat\lambda \bigl(5 p + 3 \rho\bigr) \bigl(\bar\chi + \bar{\mathcal{K}}\bigr)
+ 16 \pi \hat\lambda \rho \bigl(c_{s}^{2} \bar\omega + C \bar\sigma\bigr)
+ \bar{\mathcal{K}}' \Bigl[2 \nu \hat\lambda-  \bigl(\mu  - 2 U\bigr)\hat\delta \Bigr]
\nonumber\\
&+& \hat\delta' \Bigl[- \nu  \bar\psi + \mu  - 2 U \bigl(\bar\chi + \bar{\mathcal{K}}\bigr)\Bigr]
+ \bigl(2 U-\mu \bigr) \hat\lambda' \bar\psi  + 2 \bigl(U+\mu\bigr) \hat\lambda \bar\psi'
+ \hat\lambda \bigl(\mu ^{2} + 3 \nu ^{2} + 2 U^{2}\bigr) \bigl(\bar\chi + \bar{\mathcal{K}}\bigr)
\nonumber \\ 
&+& \bigl(3 \mu  + 2 U\bigr)\dot{\bar\chi} \hat\lambda 
+ \dot{\hat\delta} \Bigl[\bar{\mathcal{K}}' - 2 \bigl(\nu  -  W\bigr) \bigl(\bar\chi + \bar{\mathcal{K}}\bigr)\Bigr]
 + \bigl(3 \nu  + 2 W\bigr) \hat\lambda \bar\chi'
+ \bigl(3 \mu \hat\lambda  -  \hat\delta' + 8  U\hat\lambda + 2W \hat\delta\bigr)\dot{\bar{\mathcal{K}}}
\nonumber \\
&+& 2W^{2} \hat\delta \bar\psi + 2 \dot{\bar\psi} \hat\lambda \Bigl(2 \nu  + W\Bigr)
-  \dot{\hat\lambda} \bigl( \mu\bar\chi  +  \mu\bar{\mathcal{K}}  + 2  W\bar\psi\bigr)
+ 2 \nu  \Bigl[2 U \hat\lambda \bar\psi + 3 W \hat\lambda \bigl(\bar\chi + \bar{\mathcal{K}}\bigr) -   W\hat\delta \bar\psi\Bigr] \nonumber \\ 
 &-&\frac{2}{r^{2}} \Bigl[\bar{\mathcal{K}} \hat\lambda -  r^{2} U W \bigl(\hat\delta (\bar\chi + \bar{\mathcal{K}}) + \hat\lambda \bar\psi\bigr)\Bigr] + 2 \mu  \Bigl[U\hat\delta \bar\psi
+ U (\bar\chi + \bar{\mathcal{K}})\hat\lambda  + 2  (\nu+ W) \bar\psi\hat\lambda\Bigr]\biggr\},
\end{eqnarray}
\end{widetext}
where, as before, we have also excluded the $\bar l,\hat l=0, 1$ cases
by assuming vanishing $\bar \eta$ and $\hat \eta$
in order to avoid a larger expression.

On the other hand, we note that the entropy perturbation $\sigma$ is
the only matter variable appearing in the linear sources for equations
(\ref{polareq1}-\ref{polareq6}).
Its evolution follows from entropy conservation $\dot s=0$,
\begin{equation}\label{polareq10}
  \dot\sigma + (\gamma + \frac{\psi}{2})s'=
      \sum_{\bar l, \hat l}{}^{(\epsilon)}S_{\sigma},
\end{equation}
valid for $l\ge 0$, with the source terms
\begin{eqnarray}\nonumber
  {}^{(+)}S_\sigma &=& -\frac{2}{r^2}
       \E{1}{\hat l}{\hat m}{-1}{\bar l}{\bar m}{l}
       \Big\{
       (2\hat\beta\bar\lambda+\hat\delta\bar\lambda)s'-2\hat\alpha\bar\sigma
       \Big\}
       \\\nonumber
  &+& \E{0}{\hat l}{\hat m}{0}{\bar l}{\bar m}{l}
       \Big\{
       2\hat\gamma (\bar{\cal K}+\bar\chi) s'
       -(2\bar\gamma+\bar\psi)\hat\sigma'
       \\
  &+&(2\bar\eta-\bar{\cal K}-\bar\chi)\dot{\hat\sigma}
       \Big\}
       ,\\
  {}^{(-)}S_\sigma &=& -\frac{4i}{r^2} \E{1}{\hat l}{\hat m}{-1}
       {\bar l}{\bar m}{l}
       \Big\{
       \bar\alpha\hat\lambda s'-(\hat\beta+\hat\delta)\bar\sigma
       \Big\}. 
\end{eqnarray}
The second-order variable $\gamma$ in (\ref{polareq10}) must be replaced
by its expression obtained algebraically from the constraint
(\ref{polareq1}).

Therefore,
Eqs.~(\ref{polareq4}-\ref{polareq6}) and (\ref{polareq10}) form a
closed system of four evolution equations for the four unconstrained
variables $\chi$, $\cal K$, $\psi$ and $\sigma$, for $l\ge 2$.
The additional
constraint equations (\ref{polareq1}-\ref{polareq3}) provide algebraic
relations to reconstruct the matter perturbations $\gamma$, $\omega$
and $\alpha$.
The energy-momentum conservation equations provide (redundant)
evolution equations for these other fluid variables, namely,
\begin{eqnarray}\nonumber
  l\geq 0\\\label{polareq7}
      -\dot\omega-\left(1+\frac{p}{\rho}\right)
      \left(\gamma+\frac{\psi}{2}\right)'=S_\omega
      +\sum_{\bar l, \hat l}{}^{(\epsilon)}S_\omega,\\\label{polareq8}
      \left(1+\frac{p}{\rho}\right)\left(\gamma-\frac{\psi}{2}\right)\dot{}
      +c_s^2\omega'=S_\gamma +\sum_{\bar l, \hat l}
      {}^{(\epsilon)}S_\gamma,\\\nonumber
  l\geq 1\\\label{polareq9}
  -\dot\alpha=S_\alpha+\sum_{\bar l, \hat l}{}^{(\epsilon)}S_\alpha,
\end{eqnarray}
where, as before, we have gathered the non-principal part in linear
functions $S_\omega$, $S_\gamma$ and $S_\alpha$ which are given in
App.~\ref{fluidsources}. We have further defined the quadratic sources
\begin{eqnarray}
  \rho \times {}^{(\epsilon)}S_\omega & \equiv &\frac{l(l+1)}{r^2}
      {}^{(\epsilon)}{\cal E}_Au^A
      +2 U {}^{(\epsilon)}\tilde{\cal E}
      \nonumber\\
  &-&\frac{1}{r^2}\Big[r^2{}^{(\epsilon)}{\cal E}_{AB}\Big]^{|B}u^A-
      {}^{(\epsilon)}{\cal I}_Au^A,\\
  \rho \times {}^{(\epsilon)}S_\gamma& \equiv &
      \frac{l(l+1)}{r^2}{}^{(\epsilon)}{\cal E}_An^A
      +2 W {}^{(\epsilon)}\tilde{\cal E}
      \nonumber\\
  &-&\frac{1}{r^2}\Big[r^2{}^{(\epsilon)}
      {\cal E}_{AB}\Big]^{|B}n^A-{}^{(\epsilon)}{\cal I}_An^A,\quad\\
  (p+\rho) \times {}^{(\epsilon)}S_\alpha &\equiv &
      {}^{(\epsilon)}\tilde{\cal E}-
      \frac{(l-1)(l+2)}{2r^2}{}^{(\epsilon)}{\cal E}
      \nonumber\\
  &+&\frac{1}{r^2}\Big[r^2{}^{(\epsilon)}{\cal E}^A\Big]_{|A} + {}^{(\epsilon)}{\cal I}.
\end{eqnarray}

It is possible to use equations (\ref{polareq3}) and (\ref{polareq6})
to eliminate derivatives of $\psi$ in Eqs.~(\ref{polareq7})
and (\ref{polareq8}), arriving at the sound wave equation 
\begin{eqnarray}\nonumber
  l\geq 1\\\label{polareq7bar}
  -\dot\omega-\left(1+\frac{p}{\rho}\right)\gamma'=\overline{S}_\omega
      +\sum_{\bar l, \hat l}{}^{(\epsilon)}\overline{S}_\omega,
      \\\label{polareq8bar}
  \left(1+\frac{p}{\rho}\right)\dot\gamma
      +c_s^2\omega'=\overline{S}_\gamma +\sum_{\bar l, \hat l}
      {}^{(\epsilon)}\overline{S}_\gamma.\\\nonumber
\end{eqnarray}
Again, the linear sources $\overline{S}_\omega$ and $\overline{S}_\gamma$
are given in App.~\ref{fluidsources}.

\subsubsection{Polar perturbations with $l=1$}

In the special case $l=1$ there are still three polar
degrees of freedom, but the RW gauge only imposes
two gauge conditions ($H_A=0$). In consequence, the variables
${\cal K}_{AB}$ no longer serve as gauge invariants but instead
change under gauge transformations generated by the vector
$\xi_\mu dx^\mu \equiv r^2 \xi Z_a dx^a$ as
\begin{eqnarray}
  {\cal K}_{AB}\rightarrow {\cal K}_{AB} - (r^2\xi_{|A})_{|B}
      - (r^2\xi_{|B})_{|A} + \dots ,\\
  {\cal K}\rightarrow {\cal K}- 2\xi - 2 r^2 v^A\xi_{|A} + \dots ,
\end{eqnarray}
where the dots indicate source terms quadratic in the first order
perturbations.

At this point, we follow Campolattaro and Thorne \cite{CaTh70} and
fix the remaining gauge degree of freedom by demanding
${\cal K}$ to vanish. Even then we are still left with the
residual freedom of transformations under the restricted class of
functions $\xi$ such that $\xi+r^2 v^A\xi_{|A}=0$.

We have already mentioned that Eq.~(\ref{fluideq7bis}) is absent for
$l=1$, and hence we proceed differently than for $l\ge 2$.
We use equations (\ref{polareq10}), (\ref{polareq9}), (\ref{polareq7bar})
and (\ref{polareq8bar}) to evolve the
matter perturbations $\{\sigma,\alpha,\omega,\gamma\}$.
These equations contain the metric perturbations
$\{\eta,\psi,\chi\}$ but not their derivatives.
On the other hand,
we obtain the metric perturbations from
the Einstein equations (\ref{polareq1}-\ref{polareq3})
and (\ref{polareq5}-\ref{polareq6}).
With ${\cal K}=0$ from the additional
gauge freedom, these five equations contain at most first
derivatives of $\{\eta,\psi,\chi\}$. Hence, there are
five equations for six unknowns: we can obtain
$\{\dot\psi,\psi',\dot\chi,\chi'\}$
but only the following combination of $\dot\eta$ and $\eta'$,
\begin{equation}
  D\eta\equiv\frac{r^{|A}\eta_{|A}}{r^{|B}r_{|B}}
      =\frac{1}{r |v|^2}(W\eta'-U\dot\eta),
\end{equation}
where $|v|^2\equiv v^Av_A$. In consequence, we can only integrate
$\eta$ on a spatial surface that is everywhere normal to $r=const.$
surfaces, and therefore we also integrate the other scalars $\psi$
and $\chi$ in the same form,
\begin{widetext}
\begin{eqnarray}
r|v|^2D\eta&=& {\rm rhs(A12)}-\frac{1}{2}\sum_{\bar l, \hat l}
\big\{
{}^{(\epsilon)}S_{\cal K}+ (1-c_s^2) {}^{(\epsilon)}C_\omega
-2 U {}^{(\epsilon)}C_\alpha
\big\} ,\\
r|v|^2D\chi&=& {\rm rhs(A13)}+\sum_{\bar l, \hat l} \big\{2 U {}^{(\epsilon)}C_\alpha - {}^{(\epsilon)}C_\omega\big\},\\
r|v|^2D\psi&=& {\rm rhs(A14)}-\sum_{\bar l, \hat l}\big\{{}^{(\epsilon)}C_\gamma + W {}^{(\epsilon)}C_\alpha
-U{}^{(\epsilon)}S_\psi\big\},
\end{eqnarray}
\end{widetext}
where the notation rhs(X) stands for the right-hand side of Eq. (X)
of Ref. \cite{GuMa00}.

\subsubsection{The case $l=0$}

The $l=0$ perturbations are treated in a manner similar to
the case $l=1$. We evolve the matter perturbations and
then obtain the metric perturbations from the constraints.
This is a direct consequence of the absence of radiative
degrees of freedom in spherical symmetry.

For $l=0$ we still have four perturbative variables, namely
$\chi,\eta,\psi$ and ${\cal K}$, and only two gauge degrees
of freedom in the gauge-generator vector. The RW gauge does
not impose any condition for $l=0$, and hence we need to fix
those two gauge degrees in some other way.
We do this by extending the gauge used in the previous subsection
\begin{equation}
  {\cal K}=0,\qquad\psi=\frac{2UW}{U^2+W^2}(\eta-\chi).
\end{equation}
In polar-radial coordinates the second condition implies the vanishing
of the variable $\psi$.  A further feature of this case
is that the velocity perturbation $\alpha$ also vanishes.

The evolution procedure is then as follows.
The matter perturbations $\sigma$, $\omega$ and $\gamma$ are evolved
by Eqs.~(\ref{polareq10}), (\ref{polareq7}) and (\ref{polareq8}),
respectively. The latter two equations contain derivatives of
the metric perturbations
that can be removed by using the perturbed Einstein equations
(\ref{fluideq1}-\ref{fluideq3}).
The resulting equations, as well as the constraint equations for
the two non-vanishing metric perturbations $\eta$ and $\chi$
read as follows,
\begin{widetext}
\begin{eqnarray}
-\dot{\omega}-\Big(1+\frac{p}{\rho}\Big)\gamma' &=& {\rm{rhs(A15)}}
+\sum_{\bar l, \hat l}
\Big\{
{}^{(\epsilon)}S_\omega
-\frac{(p+\rho)}{2\rho|v|^2}(Uu^Au^B-Wu^An^B){}^{(\epsilon)}{\cal P}_{AB}
\Big\},
\\
\Big(1+\frac{p}{\rho}\Big)\dot{\gamma}+ c^2_s\omega' &=&
{\rm{rhs(A16)}}+\sum_{\bar l, \hat l}
\Big\{
{}^{(\epsilon)}S_\gamma
+\frac{(p+\rho)}{2\rho|v|^2}(Uu^An^B-Wn^An^B){}^{(\epsilon)}{\cal P}_{AB}
\Big\},
\\
r|v|^2D\eta &=& {\rm{rhs(A18)}}+\frac{1}{2|v|^2}\sum_{\bar l, \hat l}\big\{
(U^2+W^2)(u^Au^B+n^An^B)-4UWu^An^B
\big\}{}^{(\epsilon)}{\cal P}_{AB},
\\
r|v|^2D\chi &=&{\rm{rhs(A19)}}
+\frac{1}{|v|^2}\sum_{\bar l, \hat l}
\big\{(U^2+W^2)u^Au^B-2UWu^An^B
\big\}{}^{(\epsilon)}{\cal P}_{AB},
\end{eqnarray}
\end{widetext}
where we have again used the notation rhs(X) to refer to
the right-hand side of different formulas of Ref. \cite{GuMa00}.

\section{Perturbative matching to an exterior vacuum}
\label{sec: matching}

So far, we have studied the interior of a perfect fluid system. We now
assume that our system is a fluid star surrounded by vacuum, with both
regions, interior and exterior, separated by a surface $\Sigma$ where
the pressure vanishes. Hence, the background exterior will be
Schwarzschild.

The first part of this section describes high-order perturbative matching
across any timelike surface in a general background spacetime. The second
part will particularize to the case of a spherical fluid interior matched
to a vacuum exterior. This generalizes the first-order results of
Ref.~\cite{MaGu01} for the same scenario, and we closely follow their
notation.

\subsection{High-order matching conditions}

We describe the matching surface $\Sigma$ as the
zero level set of a smooth scalar field $f(x^\mu)$, with arbitrary smooth
continuation off the surface. The unit spacelike vector field normal to
the surface is defined as
\begin{equation}\label{defn}
  n_\mu\equiv \varphi f_{,\mu},
  \quad{\rm where}\quad \varphi\equiv (f_{,\nu}f^{,\nu})^{-1/2}.
\end{equation}
From this vector we construct the induced metric
$i_{\mu\nu}\equiv g_{\mu\nu}-n_\mu n_\nu$ and the extrinsic curvature
$e_{\mu\nu}\equiv n_{\mu;\alpha}i^\alpha{}_\nu$ of the surface $\Sigma$,
both of 
which must be continuous across $\Sigma$ to ensure a smooth matching
between the interior and exterior solutions. Before discussing high-order
perturbations of these objects we need to address two important issues
arising in the perturbative version of the matching problem:
index positioning and gauge dependence.

For a generic tensor field $T_{\mu}$ we have
\begin{equation}
\Delta[T_{\mu}]\neq g_{\mu\nu}\Delta[T^{\nu}],
\end{equation}
because in general the perturbation of the metric field does not vanish.
Imposing continuity on the perturbations of the covariant or contravariant
forms of the tensors $i$ and $e$ can therefore lead to different results,
and we must decide in advance which are the proper quantities to be used.
The first-order discussion in \cite{MaGu01} shows that we must use
perturbations of the {\em contravariant} fundamental forms. Essentially,
this is because a contravariant tensor field $T^{\mu\nu\dots}$ is
intrinsic to the surface $\Sigma$ if and only if
$T^{\mu\nu\dots}f_{,\mu}=0$ on any of its indices $\mu$.
The equivalent condition for a covariant tensor field
$T_{\mu\nu\dots}$, on the other hand,
would be $T_{\mu\nu\dots}g^{\mu\alpha} f_{,\alpha}=0$, which involves the
ambient metric and therefore introduces additional information
not intrinsic to the surface. This argument generalizes to higher-perturbative
orders, as we will later see, and so we will impose
continuity of $\Delta^n[i^{\mu\nu}]$ and $\Delta^n[e^{\mu\nu}]$
for all $n$.

Second, we need to deal with the gauge freedom arising from
the arbitrariness in our choice of mapping $\Phi$ between the perturbed
and the background spacetimes. Under a general perturbation the scalar
$f$ will also change, so that the perturbed surface will be described
by the level surfaces of $f+\Delta_\Phi[f]+...$ (a field on the
background manifold), where we have made explicit the gauge $\Phi$
relating the background and perturbed spacetimes. It is possible,
therefore, for a perturbed interior point to be located
at a coordinate position corresponding to
the background exterior. Such a situation can be handled consistently
using one-sided derivatives \cite{Mars05}. There remains, however,
the question
of the correct continuity conditions, because these conditions
may not be equivalent if expressed in different gauges. A convenient
treatment of this difficulty is possible because there exists a
privileged class of gauges $\Omega$, characterized by the condition
$\Delta^n_\Omega[f]=0$ which generalizes the first-order
{\em surface gauge} of \cite{MaGu01}.
This does not imply that the shape of the surface will
not change; indeed, the surface can be highly distorted
in the perturbed manifold. However, this kind of mappings
between perturbed and background spacetimes will identify
any point of the perturbed surface with another
point of the background surface.
In surface gauge, the matching conditions at any perturbative order
are therefore given by the continuity of
the induced metric $\Delta_\Omega^n[i^{\mu\nu}]$
and the extrinsic curvature $\Delta_\Omega^n[e^{\mu\nu}]$.

We emphasize that
surface gauge is only used to {\em define} the continuity conditions.
We can still work using any other gauge as convenient for the
interior or exterior problems. The gauge freedom is then handled by
constructing gauge-invariants associated with surface gauge,
that is, combinations of the perturbations in an arbitrary gauge whose
values coincide with the result in surface gauge. See \cite{BMM07} for
a discussion on the construction of such gauge-invariants. This can be
achieved by finding the general form of a gauge transformation from
arbitrary to surface gauge. Such transformation will
be parameterized by the gauge vectors $\{\pert{1}{\xi},...,\pert{n}{\xi}\}$
defined by solving the following equation order by order,
\begin{eqnarray}\label{eqsurfaceg}
  0=\Delta^n[f]
    +\!\!\!\sum_{m=1}^{n}\frac{n!}{(n-m)!}\!\!\!&&\!\!\sum_{(K_m)}
    \!\frac{1}{2!^{k_2}...(m!)^{k_m}k_1!...k_m!}\quad
    \nonumber\\
   &&\!\!\!\!\!\!\!\!\!\!\!\!\times {\mathcal L}_{\pert{1}{\xi}}^{k_1}...
    {\mathcal L}_{\pert{m}{\xi}}^{k_m}
    \Delta^{n-m}[f].
\end{eqnarray}
This equation is just the gauge transformation (\ref{ngaugetrans})
of the $n^{\rm th}$ order perturbation of the scalar field $f$
to surface gauge. In particular, solutions at first and second orders
are given by
\begin{eqnarray}
  \pert{1}{\xi}_\mu &=& -\Delta[f] n_{\mu},\\
  \pert{2}{\xi}_\mu &=& -\Delta^2[f] n_{\mu}+\varphi^2\Delta[f]\Delta[f]_{,\mu}.
\end{eqnarray}
These solutions fix only one of the four degrees of freedom in each
generator. There thus remain three additional degrees
of freedom which represent gauge changes among different surface gauges.

Now we can define the gauge-invariant combinations of perturbations
via an operator $\overline\Delta$ whose action on a
background tensor field $T^{\mu}$ is defined as
\begin{eqnarray}
  \overline\Delta^n[T^{\mu}] &\equiv &
      \Delta^n[T^{\mu}]
      +\!\!\sum_{m=1}^{n}\frac{n!}{(n-m)!}\!\!
      \sum_{(K_m)}\!
      \frac{1}{k_1!...k_m!}
      \nonumber\\\label{deltabar}
    &&\hspace{-0.7cm}
      \times
      \frac{1}{2!^{k_2}...(m!)^{k_m}}
      {\mathcal L}_{\pert{1}{\xi}}^{k_1}...
      {\mathcal L}_{\pert{m}{\xi}}^{k_m}
      \Delta^{n-m}[T^{\mu}],
\end{eqnarray}
where the gauge vectors $\{\pert{1}{\xi},...,\pert{n}{\xi}\}$
are those obtained by solving equation (\ref{eqsurfaceg}).
Note that the gauge generators depend implicitly on the metric
perturbations, so that this formula is non-linear and
highly nontrivial to implement.

The $\overline\Delta$ operator has been constructed explicitly so that
it obeys $\overline\Delta^n f=0$ for all $n$. Specifically, this implies
\begin{equation}\label{pertbarredn}
  \overline\Delta^n[n_\mu]=\frac{\overline\Delta^n[\varphi]}{\varphi}n_\mu.
\end{equation}
For example, at first order we obtain
\begin{equation}
  \overline\Delta[\varphi]=\frac{\varphi}{2}n^\alpha
       \left[ n^\beta \pert{1}{h}_{\alpha\beta}-2 (\varphi \Delta[f])_{,\alpha}\right].
\end{equation}
We do not display explicit second- and higher-order formulas here as they
are rather complicated and do not contribute to enlighten the discussion.

Equation (\ref{pertbarredn}) implies that perturbations of any
{\em contravariant} tensor intrinsic to the surface $\Sigma$ will also
be intrinsic to the perturbed surface. In other words, for a background
tensor $T^{\mu}$ with $T^{\mu}n_{\mu}=0$ we also have
\begin{equation}
  \overline\Delta^n[T^\mu]n_\mu =0.
\end{equation}
Note that covariant tensor fields do {\em not} share this property.

In summary, we require the barred perturbations of the contravariant
fundamental forms $\overline\Delta^n[i^{\mu\nu}]$ and
$\overline\Delta^n[e^{\mu\nu}]$ to be continuous across the surface
$\Sigma$ at any perturbative order $n$. Note that these conditions
are formulated in any gauge. In contrast to Ref.~\cite{MaGu01} we
therefore do not need to impose surface gauge.

We conclude this discussion by mentioning the alternative approach to
perturbative matching introduced by Mukohyama \cite{Muk00} and further
developed in \cite{MMV07}. Their results have been given
for first-order perturbations
and coincide with those presented here and in \cite{MaGu01}.
Mukohyama and coworkers do not
impose surface gauge and include the gauge freedom within the
matching surface $\Sigma$. Compared with our approach,
the main difference is their use of
an abstract copy of the surface $\Sigma$. The matching is
performed separately between the boundaries of the interior and exterior
spacetimes and that new surface. This procedure introduces an additional
geometric structure and gives rise to a new type of gauge invariance and,
thus, the concept of {\em double gauge-invariants}. While this can be a
convenient feature, as for example in the reduction from a $D$-dimensional
spacetime to a $(D-1)$-dimensional brane, it would complicate our
comparatively simple situation of a spherical background, where the
geometry of the matching surface is trivial.

\subsection{Matching to vacuum}

In spherical symmetry, the matching conditions can be decomposed into
tensor spherical harmonics, which allows us to extract or inject information
through the surface independently for the different harmonic components
of the metric and matter fields. We have constructed those decomposed
continuity conditions for second-order perturbations, but we will
not discuss them here, because they involve very large expressions.
Again, they coincide with the expressions given in \cite{MaGu01}
adding sources quadratic in first-order perturbations.
In App. \ref{example} we describe in detail and
explicitly how to perform the matching for the particular case of
a first-order $(l=1,m=0)$ axial and a second-order $(l=2, m=0)$
polar mode. Here we will also
mention different ways of describing the exterior perturbations.

The natural exterior frame is given in terms of the radial vector and its
orthogonal defined as
\begin{equation}\label{vacframe}
  r^A\equiv\ a\,r^{|A} \qquad t^A\equiv-\epsilon^{AB}r_B,
\end{equation}
where the normalization factor $a$ is given in terms of the Hawking mass as $a^{-1}\equiv\sqrt{1-\frac{2M}{r}}$.
In the interior of the star
these vectors are related to the fluid frame $(u^A,n^A)$
by a hyperbolic rotation.

The axial matching is simplified by the existence of the Gerlach and
Sengupta master scalar, which can be defined both in the interior and
the exterior withoug using fluid information, and so it obeys very
simple continuity conditions. The polar problem is harder because
none of the internal perturbations matches easily with the natural
variable describing vacuum perturbations, the Zerilli scalar 
\begin{eqnarray}\nonumber
\pert{n}{\Psi}&\equiv&\frac{2a^{-2}r^2}{6M+(l-1)(l+2) r}\Big(r^B \pert{n}{\mathcal K}_{AB}
-a\,r\pert{n}{\mathcal K}_{|A}\Big)r^A
\\&+&r\pert{n}{\mathcal K}.
\label{Zerilli}
\end{eqnarray}
This function obeys the following unconstrained evolution equation
\cite{BMT09},
\begin{equation} \label{Zeq}
\pert{n}{\Psi}_{|A}{}^{A} - V \pert{n}{\Psi} = \pert{n}{S},
\end{equation}
where $V$ is a potential that depends on the scalars $(r, M)$ and on the
harmonic label $l$, whereas $\pert{n}{S}$ is a source that depends on
lower-order perturbations and vanishes for linear perturbations $n=1$.
At second and higher-orders one must deal with unexpected divergencies at
asymptotic spatial infinity, which must be regularized by modifying the
definition of the Zerilli scalar with adequate lower-order terms, and
hence also modifying the form of the source of equation (\ref{Zeq}).
See \cite{BMT09} for details on the regularization procedure.

Using the gauge invariance of our perturbative functions,
we could use a particular gauge to simplify some equations.
For instance, imposing a gauge given by the conditions
$G=0$, $K=0$, and $r^Ar^BH_{AB}=0$, the Zerilli scalar is
essentially the harmonic coefficient $r^AH_A$,
\begin{equation}
\pert{n}{\Psi}^{\rm BCL} = - \frac{2a^{-1}r \,l(l+1)}{6M+(l-1)(l+2)r}
\,r^A\pert{n}{H}_A + \pert{n}{R},
\end{equation}
with $\pert{n}{R}$ being a nonlinear source depending on lower order
perturbations. The mentioned gauge is similar to the one
introduced in Ref. \cite{BCL71} ($G=0$, $K=0$, and $t^At^BH_{AB}=0$).
In fact, making use of the perturbative field equations, the linear
part of the Zerilli function turns out have the same form.

\section{Conclusions}
\label{sec: conclusions}

In this work we have presented a general formalism for
second-order perturbations of a time-dependent, self-gravitating,
spherically-symmetric perfect fluid. To this end we have extended
the formalism developed in Refs.~\cite{BMM06, BMM07}, which in turn
can be viewed as a generalization of the
Gerlach-Sengupta formalism \cite{GeSe79, GeSe80}
to higher orders. The starting point is a 2+2 decomposition
of the background spherical spacetime; the four-dimensional
manifold $\cal M$ is given by the product of a two-dimensional
Lorentzian manifold ${\cal M}^2$ and the two-sphere. This separation
enables us to fix the standard angular
coordinates $(\theta, \phi)$ on the sphere
while formulating the fields on ${\cal M}^2$ in a manifestly
covariant form. After decomposition into
tensor spherical harmonics we are left with tensors
that depend only on the coordinates on ${\cal M}^2$. We have further
defined gauge-invariant objects (for $l\geq 2$) in a manner which
identifies them with the quantities obtained in RW gauge.

An important feature of the {\em perfect} fluid in spherical symmetry is
that it defines a complete basis of vectors on the manifold ${\cal M}^2$,
given by the four-velocity of the fluid and its orthogonal. This basis has
been used to project all tensorial, background and perturbative,
quantities in such a way that one works exclusively with scalar fields.
For the case of second-order perturbations, we have explicitly decomposed
the perturbations of the energy-momentum into tensor harmonics.
By doing so, we have presented for
the first time the complete {\it matter} sources that
appear in the perturbative evolution equations.

Next, we have combined the {\it geometric} sources for the
Einstein and matter equations, as given in Ref.~\cite{BMM06},
with the matter sources derived in this article and, thus,
obtained the complete set of evolution equations.
This has been achieved by following the linear
analysis of Ref.~\cite{GuMa00} for both the polar and axial modes and
arbitrary harmonic labels $l$. Except for the polar case with
$l< 2$, we have thus obtained a combined set of evolution equations and
constraints. The metric perturbations with $l<2$
have no radiative freedom and are governed exclusively by
constraints. In this case, a solution is obtained by
first evolving the matter perturbations and then obtaining
the metric perturbations from the constraints.

We have also analyzed the problem of matching the interior perturbative
perfect fluid solution with the exterior vacuum spacetime at any order.
We have demonstrated why tensors are naturally matched
in their contravariant form. In surface gauge, defined
as the gauge where the perturbed surface is mapped
to that of the background, the perturbations of the
contravariant fundamental forms must be continuous
across the surface.
Even though this result is valid as such in
surface gauge only, the relations can be transformed to arbitrary
gauge so that practical applications are not necessarily restricted
to the use of surface gauge.

In summary, we have presented a general formalism capable of studying
coupling between any two first-order modes of a spherical time-dependent
star. In future work, we plan to apply this formalism to a variety
of stellar models including, but not restricted to, the emission of
gravitational waves by slowly rotating, collapsing stars.

\acknowledgments

This work was supported by the Spanish MICINN Project No.
FIS2008-06078-C03-03, the French A.N.R. Grant {\it LISA Science}
BLAN07-1\_201699 and by the Deutsche Forschungsgemeinschaft (DFG)
via SFB/TR7.
US acknowledges support from the Ram{\'o}n y Cajal Programme of the
Spanish Ministry of Education and Sciences (MEC),
by FCT -- Portugal through project PTDC/FIS/098025/2008 and
by grants from the Sherman Fairchild
Foundation to Caltech, by NSF grants PHY-0601459, PHY-0652995, PHY-1057238,
by loni\_numrel05
and by an allocation through the TeraGrid Advanced Support Program
under grant PHY-090003.

\appendix

\section{Polar linear sources}
\label{fluidsources}

In this appendix we give the linear sources appearing in the polar
sector of the evolution equations (\ref{polareq1})--(\ref{polareq8bar}).
These sources have been already given in Ref.~\cite{GuMa00}, but have
been recomputed in the course of this investigation, as an intermediate
step towards the construction of the second-order sources. There is full
agreement with the results of Ref.~\cite{GuMa00}.

\begin{widetext}
\begin{eqnarray}
\label{S_chi}
\nonumber
S_\chi  & = &  
- 2 \left[ 2 \nu^2 + 8 \pi \rho - {6M\over r^3} 
         - 2 U ( \mu -U )
    \right] ( \chi + {\cal K} )  
+ \frac{(l-1)(l+2)}{r^2} \chi
\\ \nonumber &+& 3 \mu \dot\chi
+ 4 ( \mu - U ) \dot {\cal K}
- ( 5 \nu - 2 W ) \chi'
- 2 [ 2 \mu \nu - 2 ( \mu - U ) W + \mu' - \dot\nu ] \psi
\\ &+&2 \eta''
- 2 ( \mu - U ) \dot\eta
+ ( 8 \nu - 6 W ) \eta'
- \left[ - 4 \nu^2 + \frac{l(l+1)+8}{r^2} + 8 \nu W 
         + 4 ( 2\mu U + U^2 - 4W^2 - 8 \pi \rho )
  \right] \eta,
\\
\label{S_k}
\nonumber S_{\cal K}
& = &
  ( 1 + c_s^2 ) U \dot\chi
+ [ 4 U + c_s^2 ( \mu + 2 U ) ] \dot {\cal K}
- W ( 1 - c_s^2 ) \chi'
- ( \nu + 2 W c_s^2 ) {\cal K}'
\\ \nonumber &-& \left[ 2 \left( \frac{1}{r^2} - W^2 \right) + 8 \pi p 
       - c_s^2 \left( \frac{l(l+1)}{r^2} + 2 U ( 2\mu + U ) - 8 \pi \rho
               \right)
  \right] ( \chi + {\cal K} )
\\ \nonumber &-& \frac{(l-1)(l+2)}{2 r^2} ( 1 + c_s^2) \chi
+ 2 [ - \mu W ( 1 - c_s^2 ) + ( \nu + W ) U ( 1 + c_s^2 ) ] \psi
+ 8\pi C \rho \sigma
\\ &-& 2 U \dot\eta + 2 W \eta'
+ \left[ \frac{l(l+1)+2}{r^2} - 6 W^2 + 16 \pi p - 2 U (2\mu+U) c_s^2
  \right] \eta,
\\
S_\psi
& = &
2 \nu ( \chi + {\cal K} ) + 2 \mu \psi + \chi' -2 \eta (\nu-W) - 2 \eta',
\\
C_\gamma
& = &
- W \dot\chi + U \chi' - ( \mu - 2 U ) {\cal K}'
+ \frac{1}{2} \left[ \frac{l(l+1)+2}{r^2} + 2 U ( 2 \mu + U )
                     - 2 W ( 2 \nu +  W ) + 8 \pi (p-\rho)
  \right] \psi
- 2 U \eta',
\\
C_\omega
\nonumber & = &
  \left[ \frac{l(l+1)}{r^2} + 2 U ( 2\mu + U ) - 8 \pi \rho
  \right] ( \chi + {\cal K} )
- \frac{(l-1)(l+2)}{2 r^2} \chi
+ 2 [ \nu U + ( \mu + U ) W ] \psi
\\ &+& U \dot\chi + ( \mu + 2 U ) \dot {\cal K}
+ W \chi' - 2 W {\cal K}'
- 2 \eta U ( 2 \mu + U ),
\\
C_\alpha
& = &
2 \mu ( \chi + {\cal K} ) + 2 \nu \psi + \dot\chi + 2 \dot {\cal K} 
- 2 \eta ( \mu + U ),
\\
S_\omega
& = &
\left( 1 + \frac{p}{\rho} \right)
\left[ - \frac{l(l+1)}{r^2} \alpha
       + \frac{\dot\chi + 3\dot {\cal K}}{2}
       + \left( \nu + 2 W - \frac{\nu}{c_s^2} \right) 
         \left( \gamma + \frac{\psi}{2} \right)
\right] 
+ ( \mu + 2 U ) \left( c_s^2 - \frac{p}{\rho} \right) \omega 
\nonumber \\ &-& C \left[ \left(\gamma+\frac{\psi}{2}\right)\frac{s'}{c_s^2}
           - \sigma (\mu+2U) 
    \right],
\\
S_\gamma
& = &
\left( 1 + \frac{p}{\rho} \right)
\left[ \frac{\chi' + {\cal K}' - 2 \eta'}{2}
       + \left[ c_s^2 (\mu + 2 U ) - \mu \right]
         \left( \gamma - \frac{\psi}{2} \right) 
\right]
- \nu \left( c_s^2 - \frac{p}{\rho} - \frac{\rho + p}{c_s^2}
               \frac{\partial c_s^2}{\partial \rho}
      \right) \omega
\nonumber \\ &-& C \sigma'
-\sigma\left[ C \left(\nu-\frac{s'}{c_s^2}
                         \frac{\partial c_s^2}{\partial s}\right)
            + s' \frac{\partial C}{\partial s}
            - \nu \left(1+\frac{p}{\rho}\right) 
              \frac{1}{c_s^2}\frac{\partial c_s^2}{\partial s}
       \right]
- \omega s' \left[ \frac{\partial c_s^2}{\partial s}
                 - C \left(1+\frac{\rho}{c_s^2}
                   \frac{\partial c_s^2}{\partial\rho}\right)
            \right],
\\
S_\alpha
& = &
- \frac{{\cal K}+\chi}{2} + \eta - c_s^2 ( \mu + 2 U ) \alpha 
+ \frac{c_s^2 \omega + C \sigma}{1+\frac{p}{\rho}}, \\
\nonumber
\bar S_\omega & = & 
\left( 1 + \frac{p}{\rho} \right)
\left[ \left( - \frac{l(l+1)}{r^2} + 8\pi(\rho+p) \right) \alpha
       + \frac{\dot {\cal K}}{2}
       + (\mu+U) \eta - \mu (\chi+{\cal K})
\right] 
+ ( \mu + 2 U ) \left( c_s^2 - \frac{p}{\rho} \right) \omega 
\\ &+& C ( \mu+2U ) \sigma
- \frac{1}{c_s^2} 
  \left[ s' C + \left(1+{p\over\rho}\right) ( \nu-2Wc_s^2 ) \right]
  \left( \gamma+\frac{\psi}{2} \right) 
+ \nu \left(1+{p\over\rho}\right) \left( \gamma-\frac{\psi}{2} \right),
\\
\nonumber
\bar S_\gamma & = &
\left( 1 + \frac{p}{\rho} \right)
\left[ \frac{{\cal K}'}{2}
       + \left( c_s^2 (\mu + 2 U ) - \mu \right)
         \left( \gamma - \frac{\psi}{2} \right) 
       - \mu \psi - \nu (\chi+{\cal K}) + (\nu-W) \eta
\right]
\\ \nonumber &-& C \sigma'
- \sigma C \left[ \nu 
                   + \frac{s'}{C} \frac{\partial C}{\partial s}
                   - \left( \frac{\nu}{C} 
                            \left( 1 + \frac{p}{\rho} \right)
                          + s'
                     \right) 
                     \frac{1}{c_s^2} \frac{\partial c_s^2}{\partial s}
              \right]
\\ &+& \omega 
  \left[ \nu \left(\frac{p}{\rho}-c_s^2\right)
       + s' \left( C - \frac{\partial c_s^2}{\partial s} \right)
       + \left[ \nu (\rho+p)+ \rho C s' \right] 
         \frac{1}{c_s^2} \frac{\partial c_s^2}{\partial\rho}
  \right] .
\end{eqnarray}
\end{widetext}

\section{Example of matching: first-order axial (l=1,m=0) and second-order polar (l=2,m=0) modes}\label{example}

The general expressions for the matching at the stellar surface are
quite large. In order to illustrate the procedure in detail, we
therefore analyze in this appendix the matching conditions for the
particular case of a first-order
$\{l=1, m=0\}$ axial mode that, by
self coupling, generates the second-order $\{l=2,m=0\}$ polar mode.
This is a particularly interesting situation since the non-radiative
first-order mode can be interpreted as a slow rotation of the star
that produces gravitational radiation through self coupling.
For identification of the perturbative order of the individual
terms, we use a left superindex $\{1\}$ on first-order perturbations.

The background junction conditions are straightforwardly deduced
from continuity of the two first fundamental forms of the surface.
The continuous quantities include the scalars $r$, $\nu$, and $W$
defined in (\ref{eq:defmunu}) and (\ref{eq:defUV}) respectively.
Derivatives of continuous quantities in the direction of the fluid
velocity $u^A$ must also be continuous. This leads to continuity
of $U$ (\ref{eq:defUV}) and, since both derivatives of $r$ are
continuous, to continuity of the Hawking mass (\ref{Hawmass}).

In our particular case, the negative pressure of the fluid
$-p$ will be interpreted as the scalar function $f$,
since the stellar surface is characterized
by $p=0$. The negative sign has been chosen such that $f$
increases with radius $r$. Furthermore the pressure must
be continuous across the surface, whereas the energy density
$\rho$ may jump there.

The first-order axial matching is simplified by the existence of the
Gerlach and Sengupta master scalar (\ref{Pidefinition}),
which can be defined both in the
interior and the exterior without using fluid information. In fact,
a first-order junction condition for our particular example is
directly obtained from the continuity of the master
scalar $\pert{1}{\Pi}$ across the surface.
Regarding the axial vector $\pert{1}{\kappa}_A$, we only have continuity
of the timelike component $\pert{1}{\kappa_A}u^A$, that has been defined as
$\pert{1}{\delta}$ in Eq.~(\ref{framevector}).
In the gauge proposed at the end of
Subsec.~\ref{axial:l=1}, $\pert{1}{\lambda}\equiv \pert{1}{\kappa}_An^A=0$,
however, the entire axial vector
$\pert{1}{\kappa}_A$ turns out to be continuous. For simplicity
and consistency with that subsection, we will employ this gauge
here. A further quantity that must be continuous is given by
the combination
\begin{equation}
\pert{1}{\Pi}'+\frac{16\pi}{r^2}\rho\pert{1}{\beta},
\end{equation}
which depends on the axial fluid perturbation $\pert{1}{\beta}$.

The second-order polar problem is more complicated because
none of the internal perturbations matches straightforwardly with the natural
variable describing vacuum perturbations, the Zerilli scalar.
Decomposing into harmonics the objects $\overline\Delta^2[i^{\mu\nu}]$ and
$\overline\Delta^2[e^{\mu\nu}]$, we find that the following second-order
polar quantities must be continuous,
\begin{eqnarray}\label{cont1}
A_1 &=& N + S, \\\label{cont2}
A_2 &=& \chi + 2(\nu+W)N - 2\eta,\\\label{cont3}
A_3 &=& {\cal K} - 2NW,
\\\label{cont4}
A_4 &=& \psi + 2 \dot{N} - 2 U N,
\\
A_5 &=& {\cal K}'- 2 \eta W +2 U \dot{N} + 2 \bigg(4\pi\rho -2U^2+W^2
\nonumber\\\label{cont5} &+&\nu W-\frac{M+4r}{r^3} \bigg) N, \\
A_6 &=& \chi' + 2\mu\psi-2\eta'+2(W+\nu)\eta -2 \ddot{N}+ 2 U \dot{N}
\nonumber\\\label{cont6}
&-&2 \left(4\pi p+2\nu^2+U^2-\nu W +\frac{M-5r}{r^3}\right) N,
\end{eqnarray}
where the quadratic source
\begin{equation}
S\equiv -\frac{1}{\sqrt{5\pi}}\pert{1}{\delta} \pert{1}{\delta} ',
\end{equation}
appears only in the first continuous combination $A_1$,
and $\eta$ is given algebraically in terms of first-order
perturbations (\ref{fluideq7bis}),
\begin{eqnarray}
\eta&=&\frac{1}{\sqrt{5\pi}}\bigg\{8 \pi(p+\rho)\pert{1}{\beta}^2
+\frac{1}{2}\left[(\nu-2 W)\pert{1}{\delta}+\pert{1}{\delta}'\right]^2
\nonumber\\&&\qquad
+\frac{\pert{1}{\delta}^2}{r^2}\bigg\}.
\end{eqnarray}
Note that some terms of the right-hand side of this
expression are continuous, e. g. $\pert{1}{\delta}^2/r^2$. Hence,
when introducing this expression in the definition of
the continuous objects (\ref{cont1}-\ref{cont6}), those
terms can be removed. We note, however, that special care is required for
the term $\eta'$ in Eq.~(\ref{cont6}), because
prime derivatives of continuous objects do not have to be continuous.
With the exception of $N$,
all perturbative objects that form part of the expressions for the
continuous objects have been defined in the main body of the
article. The remaining variable
$N$ is proportional to the gauge invariant
associated to the pressure perturbations $p$ by
\begin{equation}
-\frac{N}{\varphi}\equiv \Delta^2[p]+2{\mathcal L}_{\pert{1}{p}}\Delta [p]+
\left({\mathcal L}_{\pert{2}{p}} +{\mathcal L}^2_{\pert{1}{p}}\right)p.
\end{equation}
The pressure $p$ must not be confused with the vector $\pert{n}{p}^\mu$,
whose harmonic components are given in Ref.~\cite{BMM07}.
Last relation can be written in terms of the second-order
gauge-invariant perturbations of the energy-density
and entropy,
\begin{equation}
N\equiv -\varphi\Delta^2_{\rm GI}[p] = -\varphi\rho (c_s^2\omega+C\sigma).
\end{equation}
The subindex GI denotes the perturbation expressed in terms of
the gauge-invariant objects that, again, has a
form equivalent to the perturbation in the RW gauge $\pert{n}{p}^\mu=0$.
This last equation has no quadratic terms in first-order
energy-density and entropy perturbations because
both are polar and we assume the first-order perturbations to be
exclusively axial.
The norm of the normal vector is defined by (\ref{defn}),
\begin{equation}
\varphi =  \left(p_{,A}p^{,A}\right)^{-1/2}= -\frac{1}{p'}.
\end{equation}
The second equality holds because the pressure vanishes
at the surface at all times; therefore $\dot p=0$. The minus sign comes from
the fact that $p'<0$. On the other hand,
making use of the background Euler equation for the fluid
(\ref{Euler}), we obtain
\begin{equation}
p' =-\nu(p+\rho).
\end{equation}
Combining the last three equations, we finally obtain $N$ in terms
of the fluid variables on the surface,
\begin{equation}
N=-\frac{c_s^2\omega + C\sigma}{\nu}.
\end{equation}

As expected,
the continuity conditions (\ref{cont1}-\ref{cont6})
reduce to the first-order expressions given in \cite{MaGu01}
in the absence of the source $S$ and the component $\eta$.

\subsection{Extraction}

In the interior of the star, the relation between the natural exterior
frame $(r^A,t^A)$, defined in (\ref{vacframe}), and the fluid frame
$(u^A,n^A)$ is given by a hyperbolic rotation,
\begin{eqnarray}
r^A &=& -a\,r\,U u^A + a\,r\,W n^A,
\\
t^A &=& -a\,r\,W u^A + a\,r\,U n^A,
\end{eqnarray}
Replacing this form of the radial vector $r^A$
in the definition of the Zerilli function (\ref{Zerilli})
and rewriting it in terms of the fluid variables,
one obtains
\begin{eqnarray}
\Psi &=&\frac{2r^4}{6 M+\left(l+2\right)\left(l-1\right) r} \left\{
{U \dot{{\cal K}}_{\rm out}}
+{2 U W \psi_{\rm out}}
-{W {\cal K}_{\rm out}'}
\right.\nonumber\\
&-&\left.{2 U^2 \eta_{\rm out}}
+ {\left(U^2+W^2\right)(\chi_{\rm out}}+{\cal K}_{\rm out}) \right\} + r{\cal K}_{\rm out},
\end{eqnarray}
where primes and dots are always expressed in the fluid frame, and
the expression is evaluated just outside the surface.
Making use of the continuous quantities (\ref{cont1}-\ref{cont6}),
we arrive at the formula that gives the outside Zerilli function
in terms of the fluid inner variables,
\begin{eqnarray}\nonumber
\Psi &=&\frac{2r^4}{6 M+\left(l+2\right)\left(l-1\right) r} \big\{
{U \dot{{\cal K}}_{\rm in}}
+{2 U W \psi_{\rm in}}
-{W {\cal K}_{\rm in}'}
\\\nonumber
&-&{2 U^2 \eta_{\rm in}}+ {\left(U^2+W^2\right)(\chi_{\rm in}}+{\cal K}_{\rm in})
-8\pi r^2\rho W N_{\rm in}
\\
&-&\frac{W}{r^2}(-8+l+l^2)(S_{\rm out}-S_{\rm in})
\Big\} + r{\cal K}_{\rm in}.
\end{eqnarray}
At first order, because of the vanishing of the source $S$,
the last term in curly brackets would disappear.

\subsection{Injection}

Inside the star, the polar variables $\chi$ and ${\cal K}$ satisfy a wave
equation. Therefore, boundary conditions on the surface
of the star must be given for these variables.

The tensor ${\cal K}_{AB}$ is decomposed in Schwarzschild
coordinates as,
\begin{eqnarray}
{\cal K}_{AB}\equiv\left(
\begin{array}{cc}
H_{rr} & H_{rt}\\
H_{rt} & H_{tt}
\end{array}
\right).
\end{eqnarray}
Outside the star the polar metric perturbations
$\{H_{rr},H_{rt},H_{tt},{\cal K}\}$ can be recovered
once the Zerilli function has been determined \cite{BMT09}.
Transfer of information from the exterior to the interior
of the star is again obtained from a
hyperbolic rotation and use of the continuous quantities
(\ref{cont1}-\ref{cont6}). Specifically, using
the continuous quantities (\ref{cont2}) and
(\ref{cont3}), we get
\begin{eqnarray}
\chi_{\rm in} &=& \chi_{\rm out} -2(\nu+W)(S_{\rm out}-S_{\rm in})
\nonumber\\
&-&2(\eta_{\rm out}-\eta_{\rm in}),
\\
{\cal K}_{\rm in} &=& {\cal K}_{\rm out} + 2 W (S_{\rm out}-S_{\rm in}).
\end{eqnarray}
Applying a hyperbolic rotation results in
\begin{eqnarray}
\chi_{\rm out} &=&\frac{1}{(r-2 M)^2}
\left\{r^4U^2 H_{tt} +2r^3(r-2 M) U W H_{rt}
\nonumber\right.\\
&+&\left. r^2(r-2 M)^2 W^2H_{rr}\right\}-{\cal K}_{\rm out}.
\end{eqnarray}
Hence, we have succeeded in writing the interior variables $\{{\cal K}, \chi\}$
in terms of the exterior variables $\{{\cal K}, H_{tt}, H_{tr}, H_{rr}\}$,
which, in turn, can be obtained in terms of the Zerilli function.

\end{document}